\documentclass[romanappendices,journal,compsoc]{IEEEtran}

\usepackage[numbers]{natbib} 
\usepackage[breaklinks=true,colorlinks]{hyperref}
\usepackage{booktabs,tabularx}
\usepackage{array}
\usepackage[table]{xcolor} 
\usepackage{amsmath}
\usepackage{graphicx}
\usepackage{subfigure}
\usepackage{ragged2e} 
\usepackage{orcidlink}
\usepackage[accsupp]{axessibility} 

\usepackage[utf8]{inputenc} 
\usepackage{comment}
\usepackage[normalem]{ulem} 

\graphicspath{{Figures/}}

\begin{document}

\title{BODYinTRANSIT-Lab: A Multimodal Dataset for Investigating Sensory-Driven Transformations in Body Perception and Affective Experiences}

\author{
    Mohammad Mahdi~Dehshibi~\orcidlink{0000-0001-8112-5419}~\IEEEmembership{Senior~Member, IEEE},
    Tom\'{a}s~Mart\'{i}nez~Cort\'{e}s~\orcidlink{0000-0002-8528-5521},
    Amar~D'Adamo~\orcidlink{0009-0005-5863-692X},
    Daniel~de la Prida~\orcidlink{0000-0002-4386-5544},
    Luis-Antonio~Azpicueta-Ru\'{i}z~\orcidlink{0000-0002-2458-0914},
    Laia~Turmo~Vidal~\orcidlink{0000-0002-1769-0138},
    Marte~Roel Lesur~\orcidlink{0000-0003-2935-4868},
    Joaqu\'{i}n~R.~D\'{i}az-Dur\'{a}n~\orcidlink{0000-0002-7180-0721},
    Alexander V\"{a}ljam\"{a}e~\orcidlink{0000-0001-6071-3211},
    Angel~S\'{a}nchez~\orcidlink{0000-0003-1874-2881},
    Temitayo~Olugbade~\orcidlink{0000-0002-2838-6131},
    Nadia~Bianchi-Berthouze~\orcidlink{0000-0001-8921-0044},
    Fernando~D\'{i}az-de-Mar\'{i}a~\orcidlink{0000-0002-6437-914X},    
    and Ana~Tajadura-Jim\'{e}nez~\orcidlink{0000-0003-3166-3512}%
    \thanks{\upshape
        \textit{Manuscript received Month Day, 2026; revised Month Day, Year.} \\
        This work involved human subjects. Approval of all ethical and experimental procedures and protocols was granted by the Committee of Ethics in Research of the Universidad Carlos III de Madrid.\\
        M.M. Dehshibi is with the School of Computer Science and Creative Technologies, University of the West of England, Bristol, U.K.\\
        T. Mart\'{i}nez, A. D'Adamo, M. Roel Lesur, J. D\'{i}az and A. Tajadura-Jim\'{e}nez are with the Department of Inform\'{a}tica, Universidad Carlos III de Madrid, Legan\'{e}s, Spain. \\
        A. S\'{a}nchez is with the Department of Mathematics, Universidad Carlos III de Madrid, Legan\'{e}s, Spain. \\
        F. D\'{i}az and L.A. Azpicueta-Ru\'{i}z are with the Department of Signal Theory and Communications, Universidad Carlos III de Madrid, Legan\'{e}s, Spain. \\
        D. de la Prida is with the Architectural Acoustics Research Group, Universidad Polit\'{e}cnica de Madrid, Madrid, Spain.\\
        L. Turmo is with the Department of Media Technology and Interaction Design, KTH Royal Institute of Technology, Stockholm, Sweden.\\
        A. V\"{a}ljam\"{a}e is with the HCI group, School of Digital Technologies, Tallinn University, Estonia.\\
        T. Olugbade is with the School of Engineering \& Informatics, University of Sussex, Brighton, UK.\\
        N. Berthouze and A. Tajadura-Jim\'{e}nez are with the UCL Interaction Centre, University College London, London, UK.
    }
}

\markboth{IEEE TRANSACTIONS ON AFFECTIVE COMPUTING}%
{Dehshibi \MakeLowercase{\textit{et al.}}: BODYinTRANSIT-Lab Dataset}

\IEEEtitleabstractindextext{
\justify
\begin{abstract}
Multimodal datasets that combine synchronised behavioural and physiological signals with subjective affective and body perception measures are still rare. This limits both the understanding of embodied experience and the development of computational models that infer subjective states from sensor data. We introduce BODYinTRANSIT-Lab, a multimodal dataset that addresses this gap by capturing affective, perceptual, and behavioural responses to the \textit{Footsteps Illusion}. In this closed-loop paradigm, participants perform walking tasks while receiving real-time manipulated auditory feedback of their own footsteps intended to alter perceived body weight. The dataset comprises synchronised recordings from 101 participants recruited through systematic stratification by eating-disorder symptomatology and physical activity level. BODYinTRANSIT-Lab combines full-body kinematics with electromyography, electrocardiography, respiration, and electrodermal activity. Its annotations capture both immediate and post-trial experience, including within-trial perceived-weight ratings, post-trial assessments of body feelings and affective experience, 3D avatar body visualisations, and quantitative body maps. Psychological, behavioural, demographic, and social measures provide further information about factors that may influence individual responses. We provide a deep learning benchmark that uses walking kinematics to classify relative affective valence and perceived weight within subjects. On within-session, segment-level splits, the baseline model achieved mean balanced accuracies of 45.9\% for affective valence and 44.5\% for perceived weight. The BODYinTRANSIT-Lab dataset provides an experimentally controlled resource for cognitive neuroscience and affective computing research, with potential applications in affect-aware interactive systems, health and well-being technologies, and creative embodied experiences.
\end{abstract}

\begin{IEEEkeywords}
Multimodal Dataset, Body Perception, Affective Computing, Human-Computer Interaction, Wearable Sensors, Kinematics, Physiology, Deep Learning, Multisensory Perception, Sound, Body Illusion, Body Transformation Experiences.
\end{IEEEkeywords}}

\maketitle
\IEEEraisesectionheading{\section{Introduction}
\label{sec:introduction}}
\IEEEPARstart{O}{ur} perception of our own body shapes how we move, feel, and interact with the world~\cite{gallagher2005how}. Because this perception is continuously updated by sensory integration, it can change even when the physical body does not~\cite{Azanon2016,de_vignemont_mind_2018}. By deliberately manipulating multisensory feedback, interactive technologies can reshape mental body representations~\cite{Petkova2008, preston_illusory_2014, preston_illusory_2016, Maister_Slater_Sanchez-Vives_Tsakiris_2015, lenggenhager_video_2007} and induce Body Transformation Experiences (BTEs). BTEs involve temporary changes in how individuals perceive their bodies, including appearance, size, weight, ownership, and physical capabilities ~\cite{turmo2024body}. These transformations can alter affective experience, cognition, and behaviour, with potential implications for health and well-being~\cite{Dijkerman_Lenggenhager_2018,preston_illusory_2016,Maister_Slater_Sanchez-Vives_Tsakiris_2015}. Understanding how these experiences relate to measurable bodily responses, and why those responses differ across individuals, is therefore an important problem for affective computing.

The Footsteps Illusion offers a compelling setting for investigating this relationship during everyday movement. By modifying the frequency spectrum of self-generated footstep sounds in real time, auditory feedback can alter how heavy or light people perceive their bodies to be, alongside changes in emotional state and gait~\cite{tajadura2015light}. Walking has a dual role in this paradigm: it generates the sensory feedback that shapes bodily experience, while also providing a behavioural signal through which that experience can be investigated. This action-dependent feedback loop brings body perception, affect and movement into the same experimental setting, making the paradigm particularly suitable for examining how subjective bodily experience relates to measurable behavioural responses~\cite{tajadura-jimenez_as_2019,dadamo2024soniweight}.

However, the experience elicited by altered bodily feedback cannot be inferred from the manipulation alone. Responses to the Footsteps Illusion vary with body-related beliefs and aspirations, eating-disorder symptomatology, physical activity, and social context~\cite{tajadura-jimenez_body_2022,dadamo2024soniweight,dadamoiScience2025}. The same feedback may therefore have different perceptual and affective significance for different people. For affective computing, this raises two connected questions: how do self-reported changes in body perception and affect relate to recorded kinematic and physiological signals, and which individual characteristics are associated with variation in these responses? Addressing these questions requires an experimental record that connects the applied feedback, reported experience, movement, physiology, and participant characteristics.

Affective and gait datasets provide important foundations for modelling emotion from movement, pain-related expression, and psychological characteristics associated with walking~\cite{aung2015automatic,cosma2024psymo,zhao2025multimodal}. However, computationally investigating BTEs imposes additional requirements on the relationship between experimental design and annotation. Researchers must distinguish the sensory manipulation from the experience reported in response to it, examine within-person responses across repeated exposures, and relate these responses to characteristics that may explain individual variability~\cite{Hens2025Last, Ortigoso2025Lightweight, Dehshibi2023Pain}. Repeated ratings can capture how perceived weight varies within a trial, whereas post-trial reports capture complementary aspects of the experience. Yet existing resources were designed for different scientific objectives and do not integrate these elements within a single experimental framework.

We introduce \textbf{BODYinTRANSIT-Lab}, a multimodal dataset designed to support these investigations. It contains recordings from 101 adults, recruited with systematic stratification by eating-disorder symptomatology and physical activity level. Participants completed six counterbalanced walking trials under three auditory-feedback conditions, while full-body kinematics were synchronised with electrocardiography (ECG), electrodermal activity (EDA), electromyography (EMG), acceleration (ACC), and respiration. Subjective annotations include repeated within-trial perceived-weight ratings, post-trial body feelings and affective valence, arousal, and dominance, quantitative body maps, and avatar-based body-size estimates. Psychological, behavioural, demographic, anthropometric, and social measures provide context for investigating individual variability in these responses. BODYinTRANSIT-Lab therefore links controlled sensory manipulation with synchronised movement, physiological recordings, and complementary descriptions of bodily and affective experience. It provides data for studying BTE-based technologies intended to influence body perception, affect, and behaviour. We further describe the data-curation protocol as a basis for developing comparable multimodal datasets for the computational study of BTEs.

To demonstrate the dataset's computational usability, we introduce a sensor-aggregation deep learning architecture that combines pretrained movement representations with attention pooling, and report initial within-session, segment-level reference performance for classifying auditory condition, relative perceived body weight, and relative affective valence from walking kinematics. The benchmark demonstrates one way of studying the relationship between sensed movement and reported experience, while the multimodal recordings and participant-level information support substantially broader analyses beyond the present tasks. The contributions of this work are threefold:
\begin{enumerate}
    \item A multimodal dataset for investigating sensory-driven body transformation experiences (BTEs) that links predefined, digitally shaped auditory bodily feedback with time-aligned bodily signals, complementary subjective annotations and systematic participant characterisation.
    \item A documented experimental and data-curation protocol covering acquisition, synchronisation, preprocessing, and organisation, which provides a basis for developing comparable multimodal datasets for the computational study of BTEs.
    \item An initial computational benchmark reporting reference performance for classifying auditory condition and relative perceptual and affective labels from walking kinematics under within-session, segment-level splits.
\end{enumerate}

BODYinTRANSIT-Lab provides a resource for studying how affective and body perception experiences relate to measurable bodily responses during controlled sensory interactions, and this purpose in turn motivates the theoretical and empirical requirements that guide dataset design. The following section elaborates on these requirements and situates them within existing research on sensory-induced BTEs, affective movement datasets, and computational modelling.

\section{Background and Related Work}
\label{sec:background}

\begin{figure}[!h]
    \centering
    \includegraphics[width=1\columnwidth]{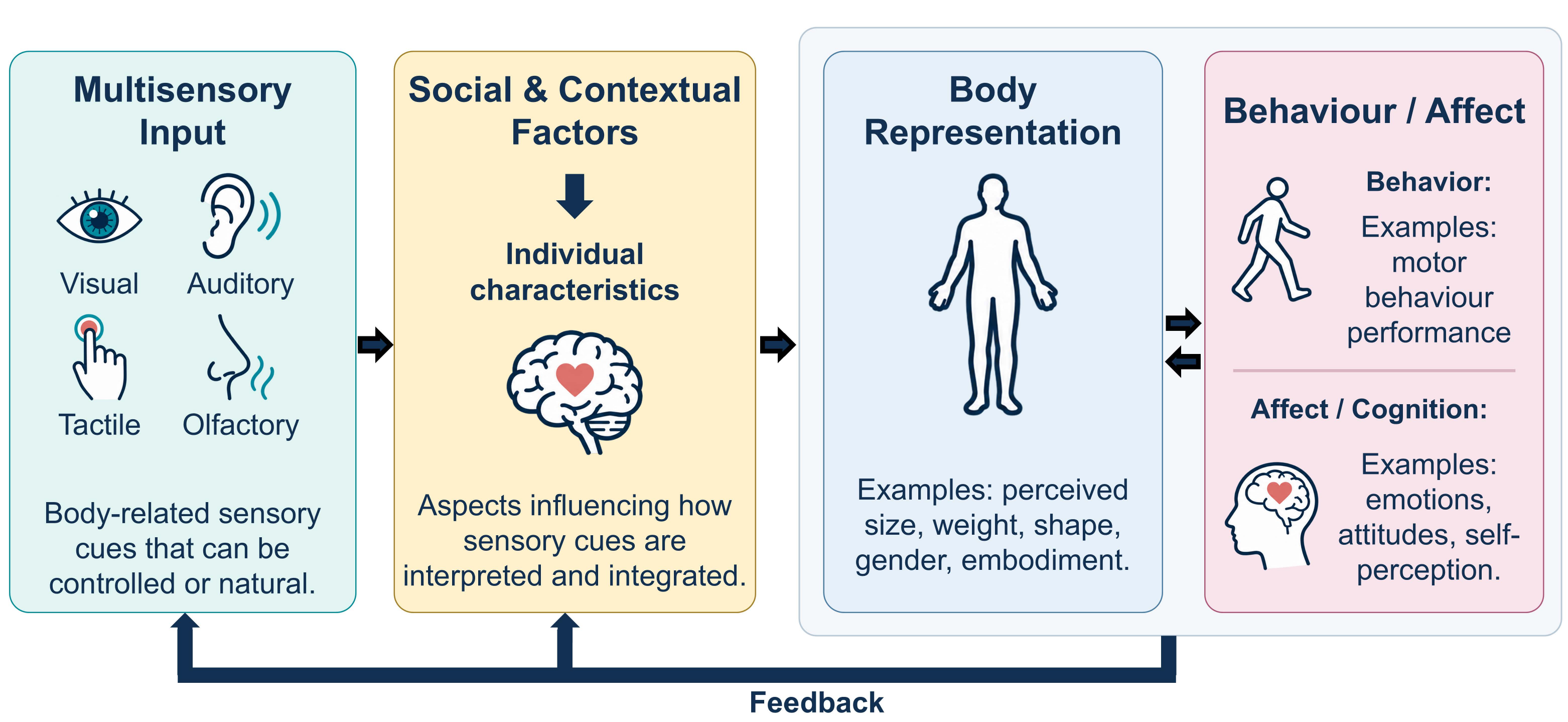}
    \label{fig:bte_framework}
    \caption{Conceptual framework for Body Transformation Experiences (BTEs), linking multisensory input, social and contextual factors, individual characteristics, body representation, and behavioural and affective responses. Arrows depict conceptual relationships and feedback, rather than causal pathways established by the present dataset.}
\end{figure}

Computational investigation of BTEs requires linking how a BTE is elicited, reported, and expressed in bodily signals. Figure~\ref{fig:bte_framework} provides a conceptual framework linking sensory inputs, individual factors, body representation, and affective and behavioural responses. These complementary observations provide the basis for computationally characterising and modelling BTEs. We review the theoretical and empirical work that motivates these requirements, followed by the datasets and modelling approaches that provide foundations for BODYinTRANSIT-Lab.

\subsection{Multisensory Body Representations, BTEs and Affective Experience}

Mental body representations integrate sensory information about the body with knowledge and beliefs about its properties~\cite{Azanon2016,gallagher2005how}. One influential account distinguishes body image, which concerns how people perceive and evaluate their bodies, from body schema, which concerns how the body is represented for posture and action~\cite{gallagher2005how,de_vignemont_mind_2018}. This distinction does not fully capture how perceptual, affective, and motor aspects of bodily experience interact~\cite{Raoul_Grosbras_2023}. A report of perceived size, a judgement of body satisfaction, and a change in movement may therefore offer complementary perspectives on a bodily transformation without measuring the same aspect of it~\cite{position_paper_preprint}. One influential account distinguishes body image, which concerns how people perceive and evaluate their bodies, from body schema, which concerns how the body is represented for posture and action~\cite{gallagher2005how,de_vignemont_mind_2018}. This distinction does not fully capture how perceptual, affective, and motor aspects of bodily experience interact~\cite{Raoul_Grosbras_2023}. A report of perceived size, a judgement of body satisfaction, and a change in movement may therefore offer complementary perspectives on a bodily transformation without measuring the same aspect of it~\cite{position_paper_preprint}.

Experimental body illusions show that multisensory stimulation can alter these representations, leading to BTEs. Botvinick and Cohen~\cite{botvinick1998rubber} demonstrated that synchronised visual and tactile stimulation can produce a sense of ownership over a rubber hand. Petkova and Ehrsson~\cite{Petkova2008} extended this work to whole-body ownership, showing that a first-person view of another body, paired with correlated bodily stimulation, can induce an experience of that body as one's own. They also measured skin conductance responses to threats directed at the artificial body, providing physiological evidence alongside subjective reports. Related experiments induced ownership of differently sized bodies and showed corresponding changes in perceived object size and distance~\cite{van2011being}. These studies established experimental methods for altering bodily self-perception by integrating sensory signals.

Further work has examined the affective significance of such changes. Preston and Ehrsson~\cite{preston_illusory_2014} induced ownership of slimmer and wider mannequin bodies and measured both perceived size and body satisfaction. Ownership of a slimmer body increased body satisfaction, whereas responses to a wider body were related to eating-disorder psychopathology. Crucianelli et al.~\cite{Crucianelli_Metcalf_Fotopoulou_Jenkinson_2013} investigated a complementary direction and found that slow, pleasant touch enhanced subjective embodiment during the rubber hand illusion without a corresponding increase in proprioceptive drift, i.e., the shift in perceived hand location toward the artificial hand. These studies illustrate a bidirectional relationship: BTEs can influence affect, while the affective qualities of stimulation can influence embodiment. They also show that different measures of embodiment need not change in parallel.

BTEs can therefore involve several interrelated perceptual, affective, behavioural, and social outcomes. Beyond affective responses, illusory body ownership has been shown to influence behaviour, social cognition, and self-identity. Embodying a child-sized body can alter perceived object size and walking patterns~\cite{van2011being}, while embodying a body of a different gender or ethnicity can shift implicit social biases and aspects of gender identity~\cite{Maister_Slater_Sanchez-Vives_Tsakiris_2015,tacikowski_perception_2020}. Because perceptual transformation alone does not determine affective meaning or behavioural expression, the embodiment literature~\cite{Roel_pos_2026} points to the need for multimodal BTE measurement that jointly captures body-perception reports, affective assessments, movement, and physiological data.

\subsection{Auditory Feedback and BTEs During Action}

Auditory approaches indicate that BTEs can arise from the sensory consequences of action. Tajadura-Jim\'{e}nez et al.~\cite{tajadura2012action} demonstrated that altered action sounds can recalibrate perceived tactile distance, consistent with the experience of having a longer arm~\cite{tajadura_action_2015}. In a related investigation of whole-body dimensions, Tajadura-Jim\'{e}nez et al.~\cite{Tajadura-Jimenez2018} manipulated the auditory and tactile feedback associated with a dropped object's impact. A longer apparent falling time led participants to report feeling taller and to behave as if their legs were longer. These studies address different bodily properties, yet both show that action-related feedback can influence estimates of the body beyond the manipulated sensory modality. In a related auditory paradigm, playback of participants' own voice and movement sounds from an allocentric perspective elicited a feeling of a self-identified presence outside the body~\cite{monologue}.

The Footsteps Illusion applies this principle to walking. Tajadura-Jim\'{e}nez et al.~\cite{tajadura2015light} altered the relative energy in different frequency bands of participants' footstep sounds and assessed perceived body weight, emotional state, and gait. Enhancing high frequencies made participants feel lighter, whereas emphasising low frequencies made them feel heavier, alongside changes in affect and gait. This established a paradigm in which an everyday full-body action generates both the manipulated sensory input and a measurable behavioural response. Clausen et al.~\cite{Clausen2021} further investigated the self-related and social significance of altered footstep sounds, examining changes in perceived gender identity and associations with gender groups. Their work broadens the outcomes relevant to auditory BTEs beyond perceived weight alone.

This paradigm distinguishes the delivered feedback from the resulting experience. The auditory condition identifies the manipulation, while subjective reports describe its perceptual and affective consequences. Preserving both enables separate analysis of information associated with the experimental manipulation and information associated with participants' reported experience.

\subsection{Individual Differences in Responses to BTEs}

The consequences of a BTE depend partly on the person experiencing it. In the Footsteps Illusion, Tajadura-Jim\'{e}nez et al.~\cite{tajadura-jimenez_as_2019} investigated the use of altered walking sounds to support physical activity and found that responses interacted with participants' body weight and masculinity--femininity aspirations. This links sensory feedback effects to the relationship between one's current and desired body. It also indicates why a transformation towards feeling lighter or heavier need not carry the same affective significance for every participant.

Eating-disorder symptomatology provides another source of variation. Tajadura-Jim\'{e}nez et al.~\cite{tajadura-jimenez_body_2022} examined the Footsteps Illusion in participants with differing symptom levels and in women with anorexia nervosa. Participants with high symptomatology or anorexia nervosa showed gait and body-visualisation patterns in the light-footstep condition that differed from those of participants with low symptomatology. The self-report measures did not show the same group differences. This result matters for both participant characterisation and annotation design, because variation in one response measure does not necessarily imply equivalent variation in the others.

D'Adamo et al.~\cite{dadamo2024soniweight} investigated several individual factors simultaneously using SoniWeight Shoes. Their study examined eating-disorder symptomatology, physical activity, body concerns, and imagery vividness in relation to perceptual, affective, physiological, and behavioural responses. Among their findings, less physically active participants reported more pronounced changes in body feelings and happiness across sound conditions, while responses in felt strength and surprise varied with body concerns. In a complementary study, D'Adamo et al.~\cite{dadamoiScience2025} examined social support networks and found stronger illusion effects among participants with smaller networks. This adds social context to the psychological and behavioural characteristics relevant to understanding response variability.

These findings motivate participant characterisation beyond demographics. Relevant dimensions include physical activity, body-related attitudes, imagery vividness, and social relationships. BODYinTRANSIT-Lab combines systematic recruitment stratification with participant-level psychological, behavioural, and social measures, allowing examination of individual variability alongside repeated within-person responses to altered feedback.

\subsection{Complementary Measures of Bodily and Affective Experience}

Subjective annotation determines which aspects of a BTE can be investigated computationally. Dimensional affect measures and body-specific reports address complementary questions. Bradley and Lang's Self-Assessment Manikin (SAM)~\cite{bradley1994measuring} provides assessments of valence, arousal, and dominance. Footsteps Illusion studies have combined these affective measures with questionnaires on bodily feelings and avatar-adjustment tasks~\cite{tajadura2015light,tajadura-jimenez_body_2022}. The latter allow participants to visually represent perceived body size. Keeping these measures distinct lets us examine whether changes in felt weight accompany changes in visualised size or affective evaluation.

\begin{table*}
    \caption{Selected affective-movement resources compared by experimental variation, recorded or released bodily data, and relevant annotations. Participant-level measures do not imply continuous annotation.}
    \label{tab:related_datasets}
    \centering
\small
\setlength{\tabcolsep}{3pt}
\renewcommand{\arraystretch}{1.05}
\renewcommand{\tabularxcolumn}[1]{m{#1}}

\begin{tabularx}{\textwidth}{@{}
    >{\raggedright\arraybackslash}m{0.16\textwidth}
    >{\raggedright\arraybackslash}m{0.26\textwidth}
    >{\raggedright\arraybackslash}m{0.22\textwidth}
    >{\raggedright\arraybackslash}X@{}}
        \toprule
        \textbf{Resource} &
        \textbf{Experimental variation} &
        \textbf{Bodily data} &
        \textbf{Annotations \& participant measures} \\
        \midrule

        Emilya~\cite{fourati2014emilya} &
        Acted emotional states across seven daily actions &
        Full-body inertial motion capture; video; audio &
        Intended emotion and third-party ratings of expressed emotion. \\

        \rowcolor[HTML]{EFEFEF}
        Li et al.~\cite{li2016emotion} &
        Walking before and after video-based induction of anger and happiness &
        Kinect-based 3D joint trajectories &
        Self-reported anger and happiness around elicitation and walking. \\

        EmoPain~\cite{aung2015automatic} &
        Rehabilitation activities under normal and increased physical demand &
        Full-body 3D motion capture; EMG; video; audio &
        Pain and anxiety reports; distress and catastrophising measures; expert annotations of pain-related expression and behaviour. \\

        \rowcolor[HTML]{EFEFEF}
        EmoPain@Home~\cite{olugbadeEmoPainATHOME2024} &
        Naturally occurring everyday activities at home &
        Wearable inertial motion capture &
        Activity labels and repeated reports of pain, worry, and movement confidence. \\

        PsyMo~\cite{cosma2024psymo} &
        Seven walking variations, including speed, carrying, and dual-task conditions &
        Video-derived silhouettes; 2D/3D skeletons; 3D body meshes &
        Personality, self-esteem, fatigue, aggression, and mental-health measures. \\

        \rowcolor[HTML]{EFEFEF}
        Zhao et al.~\cite{zhao2025multimodal} &
        Between-person variation in depression severity and personality &
        Video and audio; ankle-worn IMU gait data &
        Depression severity and personality. \\

        \textbf{BODYinTRANSIT-Lab} &
        Footsteps Illusion investigating BTEs through repeated walking with controlled changes to self-generated auditory feedback &
        Full-body inertial kinematics; ECG; EDA; EMG; respiration; backup acceleration &
        Within-trial perceived weight; post-trial body feelings and affect; body maps; avatar estimates; physical, psychological, behavioural, and social participant characteristics (Table~\ref{tab:metadata_vars}). \\

        \bottomrule
    \end{tabularx}
\end{table*}

Body maps have been used in qualitative data analysis to quantify and compare changes in body perception across experimental conditions and participant groups~\cite{Anne_Cochrane2022-ll}. They also indicate where an experience is felt. Nummenmaa et al.~\cite{Nummenmaa2014Bodily} asked participants to mark increases and decreases in bodily sensations on silhouettes and identified distinct spatial patterns associated with different emotions. Their work demonstrated how topographical self-reports can support quantitative analysis of bodily experience. In auditory BTE research, D'Adamo et al.~\cite{dadamo2024soniweight,dadamotei2026} used body maps to investigate the distribution of perceived changes across bodily regions. Such maps complement scalar ratings by preserving spatial information that a single whole-body score cannot convey.

Annotation timing introduces an additional distinction. Sharma et al.~\cite{sharma2019dataset} developed CASE to collect simultaneous, continuous valence and arousal ratings with a joystick while participants watched affective videos, along with synchronised physiological recordings. This enables the investigation of temporal variation that post-stimulus ratings alone cannot describe. In a walking paradigm, annotation must also account for the activity producing the signals of interest. BODYinTRANSIT-Lab therefore combines repeated verbal perceived-weight ratings within each trial with post-trial questionnaires, body maps, and avatar-based estimates. This design provides complementary temporal, spatial, and perceptual descriptions of experience without assuming that a single annotation method captures the entire BTE.

These approaches demonstrate the value of collecting multiple forms of subjective evidence~\cite{Roel_pos_2026}. For computational use, the annotations must also remain explicitly linked to the experimental conditions and to synchronised sensor recordings. BODYinTRANSIT-Lab links these measures within its acquisition and data-curation protocol so that researchers can examine both agreement and divergence between reported experience and bodily responses, while providing a basis for developing comparable multimodal BTE datasets.

\subsection{Affective Movement Datasets and Computational Modelling}

Research on affective body expression examines how posture and movement convey affective information~\cite{KleinsmithSurvey2013}. Multimodal affect datasets have established methods for relating bodily signals to subjective experience. DEAP~\cite{Koelstradeap} combines physiological recordings with music-video elicitation and affective ratings. AMIGOS~\cite{miranda2014amigos} adds personality, mood, and individual and group viewing contexts. Together with CASE's temporal annotation~\cite{sharma2019dataset}, these resources demonstrate the value of multimodality, participant characterisation, and context. Movement datasets extend these questions to bodily action, but differ in what varies within a participant, the bodily data they provide, and what their annotations describe. Table~\ref{tab:related_datasets} summarises selected resources along these dimensions.

Acted and affect-elicitation datasets provide complementary foundations. Fourati and Pelachaud's Emilya~\cite{fourati2014emilya} supports analysis of emotional expression across daily actions, whereas Li et al.~\cite{li2016emotion} examine walking following video-based affect elicitation. Pain-related datasets link movement to bodily experience and its affective context. EmoPain~\cite{aung2015automatic} combines rehabilitation activities with pain- and anxiety-related reports, participant-level distress and catastrophising measures, and expert annotations of pain-related expression and behaviour, supporting models such as~\cite{Ortigoso2025Lightweight,Hens2025Last, Dehshibi2023Pain} for pain and protective-behaviour detection. EmoPain@Home~\cite{olugbadeEmoPainATHOME2024} extends this line of research to everyday home activities, with repeated reports of pain, worry, and movement confidence. These resources show that intended expression, elicitation condition, bodily experience, and behavioural annotation capture distinct aspects of affective movement and should not be treated as interchangeable.

Psychologically annotated gait resources address another dimension. Cosma and Radoi's PsyMo~\cite{cosma2024psymo} combines multiple walking conditions with extensive questionnaire-derived psychological attributes, while Zhao et al.~\cite{zhao2025multimodal} examine depression severity and personality within a multimodal gait-assessment framework. These resources demonstrate that neither multimodality nor rich participant characterisation alone distinguishes BODYinTRANSIT-Lab. Within the comparison in Table~\ref{tab:related_datasets}, its distinction for BTE research lies in combining controlled changes to self-generated sensory feedback with synchronised movement and physiological recordings, repeated within-person reports of body perception and affective experience, and participant-level characteristics that can be examined as sources of between-person variability.

Advances in movement representation learning offer methods for investigating these relationships when task-specific labelled data are scarce. Yuan et al.~\cite{yuan2024self} pre-trained activity-recognition models on 700,000 person-days of unlabelled wearable data and demonstrated transfer across eight activity benchmarks. Their work provides reusable representations for sensor-based movement analysis. These representations serve as a useful computational starting point for examining whether movement contains information associated with experimentally manipulated feedback and with reported bodily or affective experience.

The reviewed literature therefore motivates an integrated resource that connects four elements, including (1) controlled action-dependent feedback, (2) synchronised bodily signals, (3) complementary reports of perceptual and affective experience, and (4) characteristics associated with individual variability. BODYinTRANSIT-Lab brings these elements together and documents their acquisition, synchronisation, curation, and organisation so that the resulting relationships can be investigated computationally and the protocol can inform the development of comparable datasets. The following section describes how these requirements informed participant recruitment, the experimental protocol, and the measures collected.

\section{BODYinTRANSIT-Lab: Study Design and Experimental Protocol}
\label{sec:dataset_design}

BODYinTRANSIT-Lab was designed to examine BTE-related responses to altered own auditory feedback during walking, while retaining information about the individuals experiencing it. The experiment combines repeated exposure to three auditory conditions with recordings of movement and physiology, within-trial perceived-weight ratings, and complementary post-trial reports, summarised in Table~\ref{tab:metadata_vars}. This structure connects the feedback delivered during a trial with the participant's reported experience and bodily signals, while participant-level measures provide context for investigating variation in these responses.

The design operates at two levels. Within participants, the auditory condition changes across repeated walking trials. Between participants, recruitment stratification and psychological, demographic, anthropometric, and social measures characterise differences relevant to BTE research. We first describe the auditory comparison, then the participant sample and experimental setup, and finally the trial sequence and subjective measures.

\begin{figure*}
    \centering
    \includegraphics[width=0.83\textwidth]{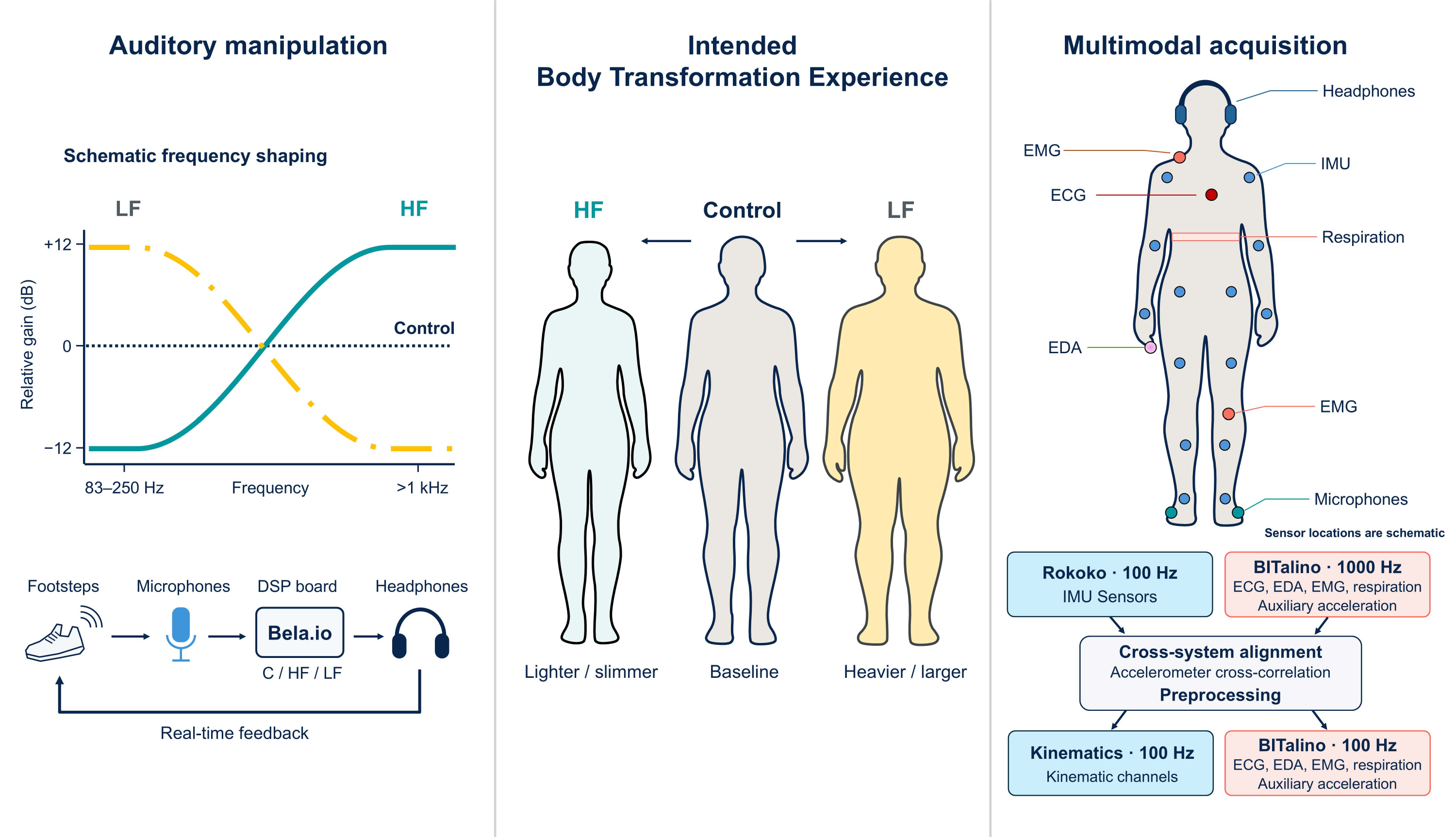}
    \caption{\textbf{Left:} real-time footstep feedback under control (C), high-frequency (HF), and low-frequency (LF) conditions; curves indicate frequency shaping. \textbf{Centre:} intended perceptual changes relative to the physical body, assessed subjectively. \textbf{Right:} kinematic and physiological acquisition, aligned by accelerometer cross-correlation, with start/stop coordinated by Node-RED~\cite{nodered}. Sensor locations are schematic.}
    \label{fig:setup}
\end{figure*}

\subsection{Auditory Conditions and Repeated-Trial Design}
\label{sec:auditory_design}

The experiment implements the Footsteps Illusion using SoniWeight Shoes~\cite{dadamo2024soniweight,DelaPrida2022}. The system records participants' footstep sounds, modifies their frequency spectra, and plays them back in real time. This preserves the action-dependent relationship between walking and the auditory feedback under investigation~\cite{menzer_feeling_2010}. Following the three-condition paradigm used in~\cite{tajadura-jimenez_body_2022}, the spectral manipulation comprised the following conditions (see Fig.~\ref{fig:setup}):
\begin{enumerate}
    \item \textbf{Control (C):} Footstep sounds were played back without spectral alteration.
    \item \textbf{High frequency (HF):} Frequencies above 1~kHz were amplified by $+12~\mathrm{dB}$, while frequencies at 83--250~Hz were attenuated by $-12~\mathrm{dB}$.
    \item \textbf{Low frequency (LF):} Frequencies at 83--250~Hz were amplified by $+12~\mathrm{dB}$, while frequencies above 1~kHz were attenuated by $-12~\mathrm{dB}$.
\end{enumerate}

Consistent with prior Footsteps Illusion studies~\cite{tajadura2015light,tajadura-jimenez_as_2019,tajadura-jimenez_body_2022}, we selected the HF and LF conditions to elicit sensations of a lighter and heavier body, respectively. These intended effects motivated the manipulation, and the subjective measures captured participants' responses. The control condition preserves playback of participants' own footsteps and serves as a baseline for comparison without spectral modification. Each participant's session comprised six main walking trials, with two repetitions of each condition presented in a counterbalanced order. This repeated-trial structure provides observations for each auditory condition within the same participant. To examine how these responses vary between individuals, the design also incorporates systematic participant characterisation.

\subsection{Participant Recruitment and Characterisation}
\label{sec:A}

BODYinTRANSIT-Lab contains recordings from 101 adults recruited through the Universidad Carlos III de Madrid participant pool and public advertisements. Eligibility required participants to be aged 18--70 years, proficient in Spanish, self-report normal hearing, be able to perform the walking tasks, and understand the study information and provide informed consent. Exclusion criteria included psychiatric or neurological diagnoses, regular use of medical or recreational psychoactive substances, disabling orthopaedic or neuromuscular conditions requiring assistance during walking, factors interfering with sensor placement, deafness or acute hearing impairment, and inability to provide informed consent.

We stratified recruitment using eating-disorder symptomatology and physical activity level, two dimensions associated with individual differences in responses to the Footsteps Illusion~\cite{tajadura-jimenez_body_2022}. Potential participants completed Spanish versions of the Eating Disorder Examination Questionnaire (EDE-Q)~\cite{edeq-esp} and the International Physical Activity Questionnaire (IPAQ)~\cite{Roman_Vinas2013-js} before the experimental session. EDE-Q scores were classified as low or high using global-score cut-offs of 1.98 for female participants and 1.09 for male participants, while IPAQ scores were separated at 3000 MET-min/week. We selected these cut-offs according to~\cite{dadamo2024soniweight}. Crossing these classifications resulted in four recruitment groups: low EDE-Q/low IPAQ ($n=20$), high EDE-Q/low IPAQ ($n=21$), low EDE-Q/high IPAQ ($n=41$), and high EDE-Q/high IPAQ ($n=19$).

\begin{figure*}[!t]
    \centering
    \includegraphics[width=0.85\textwidth]{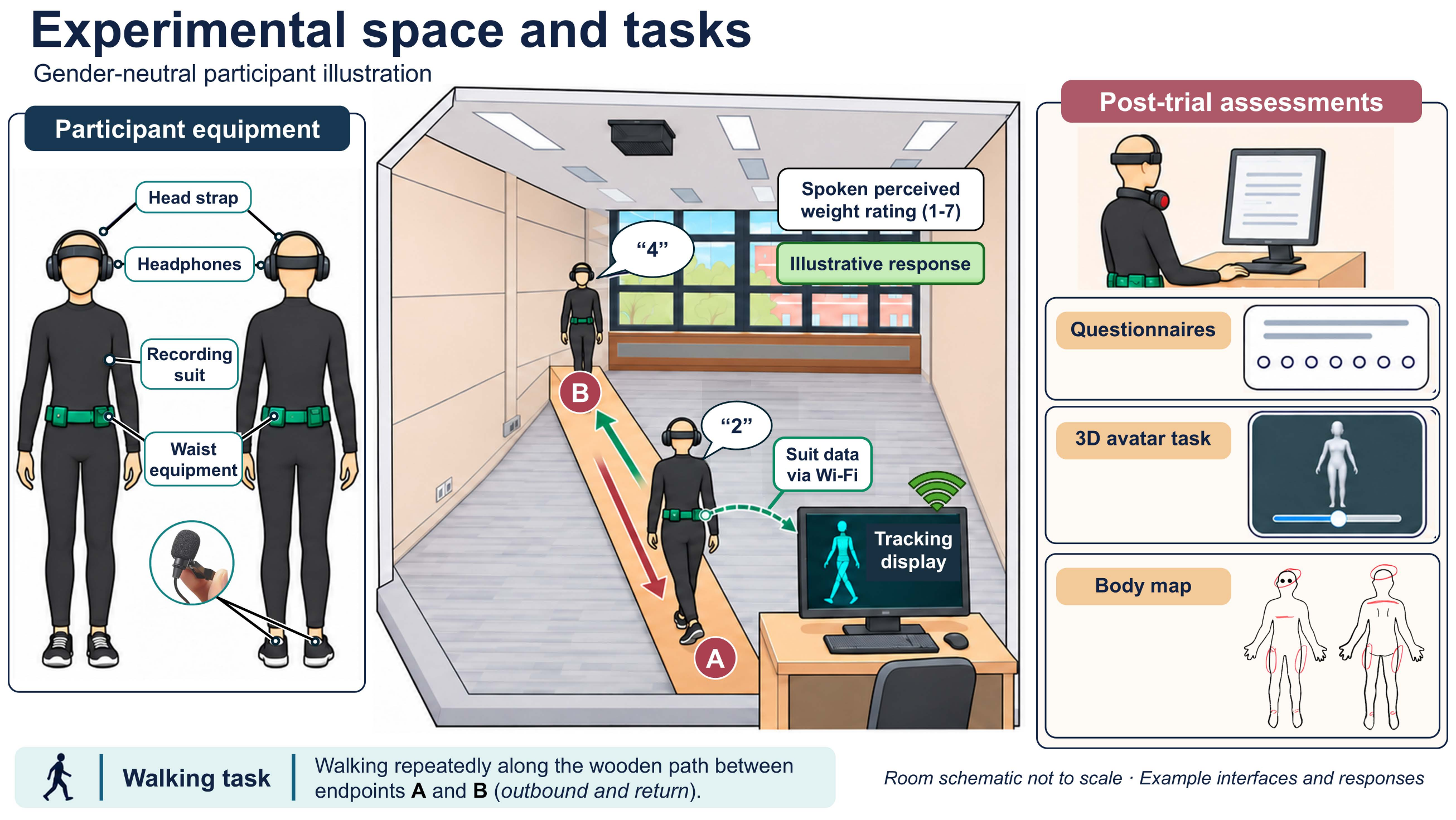}
    \caption{Experimental setting and tasks. Left: participant equipment. Centre: walking between A and B with spoken perceived-weight ratings. Right: post-trial questionnaires, avatar adjustment, and body maps. The sensor locations are shown in Fig.\ref{fig:setup}}
    \label{fig:protocol_overview}
\end{figure*}

Further measures describe aspects of participants' bodies, self-perceptions, and everyday context that are not captured by the two stratification variables. Demographic information included age, self-reported sex assigned at birth and gender identity, and highest completed education level. Sex response options were female, male, intersex, and prefer not to say, and gender identity options were man, woman, non-binary, and prefer not to say. Anthropometric and body-related information included height, actual and ideal weight, waist and hip circumferences, and shoe EU size. Body measurements also informed motion-capture suit sizing. Separate actual and ideal gender-identity ratings recorded participants' self-perceived and preferred positions on a seven-point masculinity–femininity scale.

Participants also completed Spanish versions of three questionnaires. The Multidimensional Body Self Relations Questionnaire (MBSRQ)~\cite{Roncero2015-gv} assessed attitudes towards the physical body through four scales. Betts' Questionnaire upon Mental Imagery (Betts' QMI)~\cite{Baranchok_1994-zd} assessed imagery vividness across sensory modalities, including visual, auditory, tactile, gustatory, olfactory, and organic imagery. The Social Support Questionnaire (SSQ6)~\cite{MARTINEZLOPEZ2014102} characterised the social environment using measures such as network size, sex ratio, number of social groups, and the average age of network members. These measures support the examination of associations between participant characteristics and responses to the auditory conditions. Full sample characteristics overall and by recruitment group are reported in the Supplementary Table~S1. 

The study received approval from the Committee of Ethics in Research of the Universidad Carlos III de Madrid under Application No. `CEI21\_04\_TAJADURA\_Ana' and was conducted in accordance with the Declaration of Helsinki. All participants provided written informed consent and received 20 euros for their time.

\subsection{Experimental Setting and Feedback Apparatus}
\label{sec:recording_setup}

The experimental setting supported both generating footstep sounds and recording bodily responses during walking. Sessions took place in a laboratory measuring 10.0~m long, 4.7~m wide, and 2.7~m high. Participants walked on a wooden path 7~m long and 0.8~m wide, with wood selected for its acoustic properties (see Fig.~\ref{fig:protocol_overview}). We minimised external noise during the sessions by controlling the recording environment. We positioned the microphones very close to the footwear and kept them at a consistent distance and location for all participants, ensuring a very high SNR, maintaining standardised recording conditions, and enabling the capture of the footstep sounds used for playback.

SoniWeight Shoes delivered the auditory conditions via a Bela processing board and Sennheiser HDA300 headphones. Participants wore a motion-capture suit and physiological sensors throughout the entire recording session. Figure~\ref{fig:setup} distinguishes the auditory-feedback loop from the recording systems. The acquisition specifications and sensor montage are detailed in Section~\ref{sec:sensor_data}. The hardware and software specifications are provided in Table~\ref{tab:hardware}.

\begin{table}[!h]
    \caption{Hardware and software specifications used in the experimental setup, including data acquisition and sound playback systems.}
    \label{tab:hardware}
    \centering
    \small
    \setlength{\tabcolsep}{2pt}
    \renewcommand{\arraystretch}{1.1}
    \begin{tabularx}{\columnwidth}{@{}>{\raggedright\arraybackslash}p{0.31\columnwidth}>{\raggedright\arraybackslash}X@{}}
        \toprule
        \textbf{Component} & \textbf{Details} \\
        \midrule
        Motion capture suit & Rokoko Smartsuit Pro II with 19 IMU sensors, sampled at 100~Hz. \\
        \addlinespace[2pt]
        Physiological sensing & BITalino 8-channel biosignalsplux kit; sampling rate 1000~Hz. \\
        \addlinespace[2pt]
        Auditory feedback & SoniWeight Shoes (custom)~\cite{DelaPrida2022}. \\
        \addlinespace[2pt]
        Microphones & Core Sound binaural microphone. \\
        \addlinespace[2pt]
        DSP board & Bela.io Starter Kit; average latency 1.6~ms. \\
        \addlinespace[2pt]
        Headphones & Sennheiser HDA300; closed-back, circumaural. \\
        \addlinespace[2pt]
        Control software & Bespoke JavaScript GUI implemented in Node-RED~\cite{nodered}. \\
        \addlinespace[2pt]
        Operating system & Windows 11. \\
        \addlinespace[2pt]
        Analysis software & Python 3.10 and MATLAB R2024a. \\
        \bottomrule
    \end{tabularx}
\end{table}

\subsection{Trial Sequence and Within-Trial Reports}
\label{sec:trial_sequence}

Each trial comprised a brief acclimatisation period followed by repeated walking along the path. Participants first walked in place at Point A for approximately 10~s while exposed to the assigned auditory condition, then provided an initial perceived-weight rating before beginning the walking task. They subsequently walked back and forth three times between Points A and B, 7~m apart, yielding six traversals per trial. Upon reaching each endpoint, participants verbally rated their perceived body weight on a seven-point scale from 1 (light) to 7 (heavy), providing additional ratings, for a total of seven ratings per trial. Figure~\ref{fig:protocol_overview} illustrates the walking directions and spoken ratings at the endpoints.

These repeated reports captured perceived weight at successive time points within each walking trial. Immediately following each trial, participants completed the post-trial assessments described below. Across a complete six-trial session, the protocol thus produced 42 within-trial perceived-weight ratings and six sets of post-trial assessments, together with continuous kinematic and physiological recordings. These two annotation timescales were complementary: repeated ratings captured variation in felt weight during task performance, whereas post-trial measures characterised additional aspects of the experience.

\subsection{Post-Trial Measures of Bodily and Affective Experience}
\label{sec:E}

Post-trial annotation combined scalar ratings, avatar adjustments, and body maps to capture aspects of the experience that the within-trial weight rating alone cannot describe. The Body Feelings Questionnaire, adapted from~\cite{tajadura2015light,tajadura-jimenez_as_2019}, collected seven-point ratings of bodily sensations, including perceived weight. The Self-Assessment Manikin (SAM)~\cite{bradley1994measuring} provided nine-point ratings of valence, arousal, and dominance. These measures retain the distinction between how the body feels and the affective experience accompanying that feeling. The post-trial perceived-weight rating was separate from the seven verbal weight ratings collected during walking. After each trial, participants rated how light or heavy they felt on the same seven-point (1--7) scale.

To obtain a visual estimate of felt body size, participants adjusted the weight of a three-dimensional avatar using the Body Visualizer tool~\cite{MPI-IS:BodyVisualizer} until it matched their felt body size. The resulting avatar weight, expressed in kilograms, provides an alternative measure of perceived body size, consistent with its use in previous studies~\cite{tajadura2015light,tajadura-jimenez_as_2019,Tajadura-Jimenez2017CRPS,tajadura-jimenez_body_2022}. In addition to this scalar measure, participants freely marked locations where they experienced changes in bodily sensation or perception after each trial (see Fig.~\ref{fig:protocol_overview}, right panel). These drawings were subsequently coded by region. This spatial reporting approach complements the scalar and avatar-based measures. Its conversion into quantitative data is described in Section~\ref{sec:annotation_conversion}.

The protocol therefore produces several descriptions of the same trial: the auditory condition, bodily recordings, repeated perceived-weight reports, and post-trial assessments of bodily and affective experience. The following section describes how the recordings and annotations were acquired and prepared as components of the BODYinTRANSIT-Lab resource.

\begin{table*}
    \caption{Kinematic channels exported by the Rokoko Smartsuit Pro II. Segment linear kinematics comprise position, velocity, and acceleration along the $X$, $Y$, and $Z$ axes. Each joint-angular group comprises angle, angular velocity, and angular acceleration. \textsuperscript{\ddag} 17 body segments include the head, chest, pelvis, shoulders, upper arms, forearms, hands, upper legs, lower legs, and feet.}
    \label{tab:kinematic_data}
    \centering
    \small
    \setlength{\tabcolsep}{2pt}
    \renewcommand{\arraystretch}{1.1}
    \begin{tabularx}{\textwidth}{@{}
        >{\raggedright\arraybackslash}X
        >{\raggedright\arraybackslash}X
        >{\centering\arraybackslash}m{0.10\textwidth}
        >{\centering\arraybackslash}m{0.10\textwidth}@{}}
        \toprule
        \textbf{Kinematic quantity} & \textbf{Anatomical locations} & \textbf{Groups} & \textbf{Channels} \\
        \midrule
        \rowcolor[HTML]{EFEFEF}
        \multicolumn{4}{@{}l}{\textbf{Segment linear kinematics}} \\
        Position, velocity, acceleration $(X,Y,Z)$ & 17 body segments\textsuperscript{\ddag} & 17 each & \textbf{153} \\
        \midrule
        \rowcolor[HTML]{EFEFEF}
        \multicolumn{4}{@{}l}{\textbf{Joint angular kinematics}} \\
        Flexion & Hips, knees, shoulders, elbows, wrists, neck & 11 & 33 \\
        Abduction & Wrists, elbows, shoulders, hips & 6 & 18 \\
        External rotation & Shoulders, hips, knees & 6 & 18 \\
        Pronation & Wrists, elbows & 4 & 12 \\
        Adduction & Knees, hips & 4 & 12 \\
        Dorsiflexion, inversion, internal rotation & Ankles & 2 each & 18 \\
        Extension, lateral flexion/rotation, axial rotation & Thorax, pelvis & 2 each & 18 \\
        Protraction, medial rotation, posterior tilt & Scapulae & 2 each & 18 \\
        Leftward tilt, Rightward rotation & Neck & 1 each & 6 \\
        \bottomrule
        \multicolumn{3}{r}{\textit{Angular subtotal}} & \textbf{153} \\
        \multicolumn{3}{r}{\textbf{Total kinematic channels}} & \textbf{306}
    \end{tabularx}
\end{table*}

\section{Multimodal Data Acquisition, Curation, and Dataset Structure}
\label{sec:data_curation}

The experimental protocol defines the conditions and observations needed to investigate BTEs. Their computational use also depends on how signals from separate acquisition systems are aligned, how subjective reports are represented, and how the resulting data are organised. This section follows these stages from sensor acquisition through signal and annotation preparation to the structure of the BODYinTRANSIT-Lab resource. The documented workflow provides a basis for developing comparable multimodal datasets.

\subsection{Kinematic and Physiological Acquisition}
\label{sec:sensor_data}

Kinematic signals were acquired with the Rokoko Smartsuit Pro II~\cite{rokokosmartsuit} at 100~Hz, using the configuration specified in the Rokoko software~\cite{rokokosmartsuit}. The system provides kinematics for 17 anatomical segments, including three-dimensional position, velocity, and acceleration, as well as joint angles and angular derivatives. The resulting representation contains 306 channels, comprising 153 segment-level linear kinematic channels and 153 joint-angular channels. Segment positions and joint angles are reported in the suit's body-fixed anatomical coordinate system, with segment axes aligned with the corresponding anatomical directions and joint angles defined relative to adjacent segments, following Rokoko's standard conventions. Table~\ref{tab:kinematic_data} summarises how these channels are organised by anatomical location and measurement type.

We acquired physiological signals using the BITalino system at 1000~Hz. EMG was recorded from the cervical trapezius and left vastus medialis oblique, ECG from the chest, EDA from the non-dominant hand, and respiration using a chest belt. The system also included a 3-axis ACC. After recording, we identified intermittent sensor failures affecting individual axes, so each recording contains one, two, or three acceleration axes, depending on the trial. The released dataset retains this signal as backup rather than as a complete 3-axis modality. 

These two systems record movement and physiological data during the walking trials, with their acquisition specifications summarised in Table~\ref{tab:hardware}.

\subsection{Synchronisation and Signal Preparation}
\label{sec:signal_preparation}

Recording across the two systems required both acquisition coordination and subsequent temporal alignment. A bespoke JavaScript interface, implemented in Node-RED~\cite{nodered}, sent simultaneous start and stop commands. We aligned streams using cross-correlation of accelerometer signals recorded during the initial walking-in-place period. Thus, the acclimatisation activity described in Section~\ref{sec:trial_sequence} also supplied a shared movement signal for alignment. The synchronisation code is available through~\cite{DadamoRaramuriSyncSoftware}.

Signal preparation was modality-specific, reflecting the spectral characteristics and artefacts of each physiological recording, as summarised in Table~\ref{tab:signal_preprocessing}. The preprocessing pipeline included band-pass or low-pass filtering, as appropriate, and Hampel filtering to remove transient outliers. EMG was additionally filtered with a 49--51~Hz Butterworth notch filter and full-wave rectified. BITalino recordings were processed at the original 1000~Hz sampling rate using the modality-specific procedures, then downsampled to 100~Hz. We applied no additional filtering to the Rokoko recordings. These steps describe preparation of the acquired streams. The original acquisition rates differ from the sampling rates of the processed signals, and we provide both raw and preprocessed sensor recordings.

\begin{table*}
    \caption{Preprocessing of BITalino signals. Filters were applied using zero-phase forward--backward filtering. Hampel filtering removed transient outliers using fixed or signal-dependent window lengths.}
    \label{tab:signal_preprocessing}
    \centering
    \small
    \setlength{\tabcolsep}{2pt}
    \renewcommand{\arraystretch}{1.12}
    \renewcommand{\tabularxcolumn}[1]{m{#1}}

    \begin{tabularx}{\textwidth}{@{}
        >{\raggedright\arraybackslash}m{0.17\textwidth}
        >{\raggedright\arraybackslash}m{0.21\textwidth}
        >{\raggedright\arraybackslash}m{0.25\textwidth}
        >{\raggedright\arraybackslash}X@{}}
        \toprule
        \textbf{Signal} & \textbf{Filtering} & \textbf{Filter parameters} & \textbf{Subsequent processing} \\
        \midrule
        \rowcolor[HTML]{EFEFEF}
        Respiration & Butterworth IIR band-pass & Order 2; 0.1--0.35~Hz & Adaptive Hampel filtering with window length derived from the detected breathing rate; 3-SD threshold. \\
        \addlinespace[2pt]
        EMG & FIR band-pass; Butterworth IIR notch & Band-pass order 20, 20--450~Hz; notch order 2, 49--51~Hz & Hampel window 10 samples; 3-SD threshold; full-wave rectification. \\
        \addlinespace[2pt]
        \rowcolor[HTML]{EFEFEF}
        ECG & FIR band-pass & Order $\lfloor 0.3f_s \rfloor$; 3--45~Hz & Hampel window 50 samples; 3-SD threshold. \\
        \addlinespace[2pt]
        EDA & Butterworth IIR low-pass & Order 4; 5~Hz & Hampel window 150 samples; 3-SD threshold. \\
        \addlinespace[2pt]
        \rowcolor[HTML]{EFEFEF}
        Backup acceleration\textsuperscript{\ddag} & Butterworth IIR band-pass & Order 2; 0.5--10~Hz & Hann window followed by adaptive Hampel filtering with a window length derived from the dominant frequency. \\
        \bottomrule
    \end{tabularx}
    \begin{flushleft}
        \small
        \textsuperscript{\ddag}The dominant frequency was estimated from the Hann-windowed signal and used to determine the Hampel-window length; Hampel filtering was then applied to the same windowed signal.
    \end{flushleft}
\end{table*}

\subsection{Subjective Annotation Conversion}
\label{sec:annotation_conversion}

Subjective measures retain distinct temporal and representational forms. Verbal perceived-weight ratings provide repeated observations within each trial, whereas body-feeling and SAM scores describe the post-trial assessment. The avatar-adjustment task contributes the selected avatar weight in kilograms. Preserving these distinctions ensures that the reports are interpreted according to the manner and timing of their elicitation, rather than as interchangeable descriptions of experience.

The free-form paper body maps required a subsequent coding step. We visually inspected the completed drawings and manually coded them using a 136-region representation covering the anterior and posterior body surfaces, following the method described in~\cite{dadamo2024soniweight,dadamotei2026}. We assigned each region a value from 0 to 1 based on the proportion of its area the participant marked, with 0 indicating no marking in that region. This representation retains information on the spatial distribution of reported bodily changes, along with scalar ratings and avatar estimates.

\subsection{Dataset Organisation and Data Inventory}
\label{sec:dataset_structure}

BODYinTRANSIT-Lab contains recordings from 101 participants, totalling approximately 16 hours of data and about 15~GB. The database is organised by participant. Each participant directory contains separate subfolders for BITalino, Rokoko, and Label. Trial-specific CSV filenames encode the participant identifier, auditory condition, and repetition, linking the physiological, kinematic, and trial-level annotation files across the database. The manually coded body-map annotations are stored separately in CSV files and retain the corresponding participant and trial identifiers. 

The data inventory in Table~\ref{tab:metadata_vars} separates participant-level characteristics, experimental identifiers, time-series recordings, and subjective reports. Participant identifiers, trial indices, and auditory-condition codes provide context for interpreting the observations. Distinguishing participant-level information from repeated trial-level measurements supports analysis of between- and within-individual variability. 

\begin{table*}
    \caption{BODYinTRANSIT-Lab data inventory, grouped by participant characteristics, experimental identifiers and auditory manipulation, sensor recordings, and subjective responses. The table distinguishes each entry by its level and collection timing. Detailed acquisition specifications are given in Table~\ref{tab:hardware}, and kinematic channels are listed in Table~\ref{tab:kinematic_data}.}
    \label{tab:metadata_vars}
    \centering
    \small
    \setlength{\tabcolsep}{2pt}
    \renewcommand{\arraystretch}{1.1}
    \renewcommand{\tabularxcolumn}[1]{m{#1}}
    \begin{tabularx}{\textwidth}{@{}>{\raggedright\arraybackslash}m{0.18\textwidth}>{\raggedright\arraybackslash}m{0.33\textwidth}>{\raggedright\arraybackslash}m{0.19\textwidth}>{\raggedright\arraybackslash}X@{}}
        \toprule
        \textbf{Component} & \textbf{Measures or identifiers} & \textbf{Level and timing} & \textbf{Representation} \\
        \midrule
        \rowcolor[HTML]{EFEFEF}
        \multicolumn{4}{@{}l}{\textbf{Participant characteristics and identifier}} \\
        \addlinespace[2pt]
        Participant identifier & Unique participant ID & Participant & Integer identifier \\
        \addlinespace[2pt]
        Demographic measures & Age, sex, gender, education & Participant; before session & Numeric or categorical values \\
        \addlinespace[2pt]
        Anthropometric measures & Height, waist, hip circumferences (cm); weight (kg); shoe size (EU) & Participant; before session & Numeric values in the stated units \\
        \addlinespace[2pt]
        Body-related self-perceptions & Actual and ideal gender-identity ratings; ideal weight (kg) & Participant; before session & Ordinal masculinity--femininity ratings, 1--7; numeric weight \\
        \addlinespace[2pt]
        Psychological and behavioural measures & EDE-Q, IPAQ, MBSRQ, Betts' QMI & Participant; before session & Questionnaire scores \\
        \addlinespace[2pt]
        Social measures & SSQ6 & Participant; before session & Numeric network descriptors \\
        \midrule
        \rowcolor[HTML]{EFEFEF}
        \multicolumn{4}{@{}l}{\textbf{Experimental identifiers and auditory manipulation}} \\
        \addlinespace[2pt]
        Auditory condition and trial index & Control (C), high frequency (HF), low frequency (LF); trial identifier & Trial & Condition code; index 1--6 \\
        \midrule
        \rowcolor[HTML]{EFEFEF}
        \multicolumn{4}{@{}l}{\textbf{Sensor recordings}} \\
        \addlinespace[2pt]
        Kinematic recordings & Segment position, velocity, acceleration; joint angles and angular derivatives & Time series; during trial & 306 numeric channels \\
        \addlinespace[2pt]
        Physiological recordings & ECG, EDA, EMG, respiration & Time series; during trial & Numeric signal channels \\
        \addlinespace[2pt]
        Backup acceleration & Available accelerometer axes & Time series; during trial & One, two, or three numeric channels, depending on the trial \\
        \midrule
        \rowcolor[HTML]{EFEFEF}
        \multicolumn{4}{@{}l}{\textbf{Subjective responses}} \\
        \addlinespace[2pt]
        Within-trial perceived weight & Seven verbal reports per trial & Trial; during trial & Ordinal ratings, 1--7 \\
        \addlinespace[2pt]
        Post-trial body feelings & Body Feelings Questionnaire, including perceived weight & Trial; immediately after walking & Ordinal ratings, 1--7 \\
        \addlinespace[2pt]
        Post-trial affect & SAM valence, arousal, dominance & Trial; immediately after walking & Ordinal ratings, 1--9 \\
        \addlinespace[2pt]
        Avatar-based estimate & Adjusted avatar weight representing felt body size & Trial; immediately after walking & Numeric value (kg) \\
        \addlinespace[2pt]
        Body maps & Regional extent of reported changes in bodily sensation or perception & Trial; immediately after walking & 136 regional occupancy values, 0--1; 0 denotes no marking \\
        \bottomrule
    \end{tabularx}
\end{table*}

This organisation brings controlled feedback, participant characteristics, bodily signals, and reported experiences into a common resource. The next section defines a computational benchmark using selected kinematic recordings and subjective measures, and describes its task-specific transformations as part of that analysis.

\section{Computational Benchmark}
\label{sec:methodology}

The computational benchmark examines whether walking kinematics contain information associated with the auditory manipulation and the bodily and affective experiences reported in response to it. These relationships provide an initial reference performance for evaluating the computational use of BODYinTRANSIT-Lab. We selected auditory condition, relative perceived weight, and relative affective valence as classification targets to examine the delivered manipulation, the body-perceptual change targeted by the Footsteps Illusion, and the reported affective experience. Auditory-condition labels identify the delivered feedback (C, HF, or LF), whereas the subjective labels derive from the post-trial seven-point perceived-weight rating and nine-point SAM valence rating. The physiological recordings and participant-level measures remain available for broader analyses beyond this kinematic benchmark.

For the subjective tasks, the comparison of interest is each participant's reported experience relative to their control condition. We therefore constructed labels separately for each participant and task, using the mean of the two control-trial ratings as the participant-specific reference. Both control trials were assigned label 0 by definition. For the four manipulated-feedback trials (two HF and two LF), ratings above this reference received label $+1$, ratings below it received label $-1$, and ratings equal to it received label 0. Since preliminary inspection indicated that many participants used a narrow portion of the rating scale, we did not apply a tolerance margin. Under this definition, class 0 includes the control trials regardless of their individual ratings and does not denote an independently measured absence of a BTE. Each walking segment, corresponding to one traversal, inherits the auditory condition and the relevant subjective label of its parent trial. The subjective targets therefore describe post-trial reports rather than momentary measurements of experience at each input window. The seven within-trial perceived-weight ratings, avatar estimates, and body maps described in Sections~\ref{sec:trial_sequence} and~\ref{sec:E} provide complementary annotations but were not used to construct these targets.

Individual differences in movement and subjective scale usage motivate a within-participant evaluation, with a separate model trained for each participant and task. The benchmark evaluates prediction on held-out walking segments from the same recorded session. Because segments from a single trial may appear in different subsets, the benchmark is limited to prediction within recorded sessions rather than generalisation to unseen trials, sessions, or participants. Of the 101 participants, 99 completed all six trials, and 98 also provided the questionnaire data required for the subjective tasks. Excluding participants whose labels comprised only one class left 92 participants for relative perceived weight and 83 for relative affective valence. These eligibility criteria apply to the benchmark and do not alter the dataset's composition of 101 participants.

\subsection{Model Architecture}
\label{sec:rosa_architecture}

\begin{figure}[!h]
    \centering
    \includegraphics[width=0.7\columnwidth]{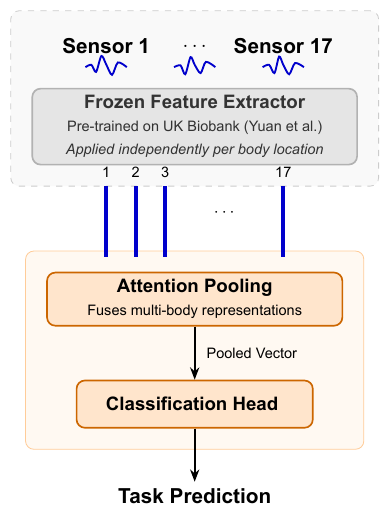}
    
    \caption{The network architecture used for the computational benchmark.}
    \label{fig:model_architecture}
\end{figure}

The proposed architecture, depicted in Figure ~\ref{fig:model_architecture}, combines the frozen movement feature extractor of the multi-task self-supervised model proposed by Yuan et al.~\cite{yuan2024self} with newly trained attention pooling and a classification head. The architecture addresses two design considerations: learning from limited participant-specific data and combining information from multiple body locations. We select the five-second checkpoint pretrained on wrist accelerometry from 100,000 UK Biobank participants~\cite{doherty2017large} and replace its original classifier with the pooling and classification components described below. Only these new components are trained for each participant and task.

The shared frozen feature extractor processes each of the 17 triaxial acceleration streams independently, producing a 512-dimensional representation for each body location. A shared linear projection scores each feature vector, and a softmax across the 17 body locations converts these scores into input-dependent attention weights. An attention-pooling layer uses these weights to combine the 17 feature vectors into a single 512-dimensional vector. This mechanism is intended to accommodate differences in signal informativeness across body locations, including those associated with the artefact masking described in Section~\ref{sec:benchmark_details}.

The classification head comprises a linear projection to 256 dimensions, a ReLU activation, and a final linear layer producing three class logits. The frozen extractor provides a common movement representation, while the trainable pooling and classification layers determine how information from the different locations contributes to the prediction.

\subsection{Benchmark Inputs and Evaluation}
\label{sec:benchmark_details}

The benchmark uses the three acceleration components from each of the 17 body segments, giving 51 input channels from the 306 kinematic channels described in Section~\ref{sec:sensor_data}. Each observation contains 150 samples per channel, matching the pretrained model's input size. We downsampled the 100~Hz recordings by a factor of three using a normalised triangular kernel $[0.25,~0.5,~0.25]$. This downsampling was applied independently to each channel using grouped one-dimensional convolution with stride 3 and one-sample padding. At the resulting rate of approximately 33.3~Hz, 150 samples span 4.5~s. The pretrained model used 5~s windows sampled at 30~Hz. These transformations prepare inputs for the benchmark and are separate from the dataset-curation procedures in Section~\ref{sec:data_curation}.

We applied artefact masking separately to each acceleration axis to limit the influence of unusually low or high signal variation within a window. Masking uses each axis's dynamic range within a window, defined as the difference between its max and min values, relative to an axis-specific reference interval. The interval spans the 5th to 95th percentiles of dynamic ranges from five participants' recordings selected by visual inspection for signal quality. These participants were also included in the benchmark, so the reference recordings were not independent of the evaluation sample. During training and evaluation, an axis with a dynamic range outside its reference interval was set to zero. Min--max normalisation to $[-1,1]$ then places each axis within each window on a common scale while retaining the temporal pattern of non-constant signals.

A complete session provides 36 walking segments after acclimatisation, corresponding to six traversals in each of six trials. For each participant and random seed, we partitioned these segments, rather than individual windows, so that every window from a given segment belonged to the same subset. We used Monte Carlo cross-validation with 10 random seeds and a two-stage split into training, validation, and test subsets at target proportions of 60\%, 20\%, and 20\%. For stratification, we considered auditory condition and task label jointly. However, when the available observations did not permit stratification, the split was random. The subjective labels were defined before partitioning, using the mean of both control-trial ratings as the participant-specific reference. Within each subset, we randomly sampled walking segments and window starting positions for training, validation, and testing. To compute the reported metrics, we used individual window predictions without aggregation into segment- or trial-level decisions.

For each participant, task, and random seed, the trainable components of the benchmark model were optimised using a weighted cross-entropy loss. Class weights were computed as the inverse of the class frequencies in the training subset for each run, to increase the contribution of less frequent labels to the loss. Models were optimised with Adam using a learning rate of $1\times10^{-4}$, a batch size of 64, and a maximum of 100 epochs. Training stopped when validation loss did not improve for 10 consecutive epochs or when validation balanced accuracy reached 1.0. The checkpoint with the lowest validation loss was retained for evaluation on the test set.

Performance was evaluated using balanced accuracy (BA), macro F1, and overall accuracy (Acc). BA averages class-specific recall, whereas macro F1 averages the harmonic mean of precision and recall within each class. These metrics are calculated using Eq.~\eqref{eq:metrics}, where $K$ denotes the number of evaluated classes and $\mathrm{TP}_c$, $\mathrm{FP}_c$, and $\mathrm{FN}_c$ denote true positives, false positives, and false negatives for class $c$, respectively.

\begin{align}
    \label{eq:metrics}
    \mathrm{BA} &=
    \frac{1}{K}\sum_{c=1}^{K}
    \frac{\mathrm{TP}_c}{\mathrm{TP}_c+\mathrm{FN}_c},
    \nonumber\\
    \mathrm{F1} &=
    \frac{1}{K}\sum_{c=1}^{K}
    \frac{2\mathrm{TP}_c}
    {2\mathrm{TP}_c+\mathrm{FP}_c+\mathrm{FN}_c}.
\end{align}

\begin{figure}[!h]
    \centering
    \includegraphics[width=\linewidth]{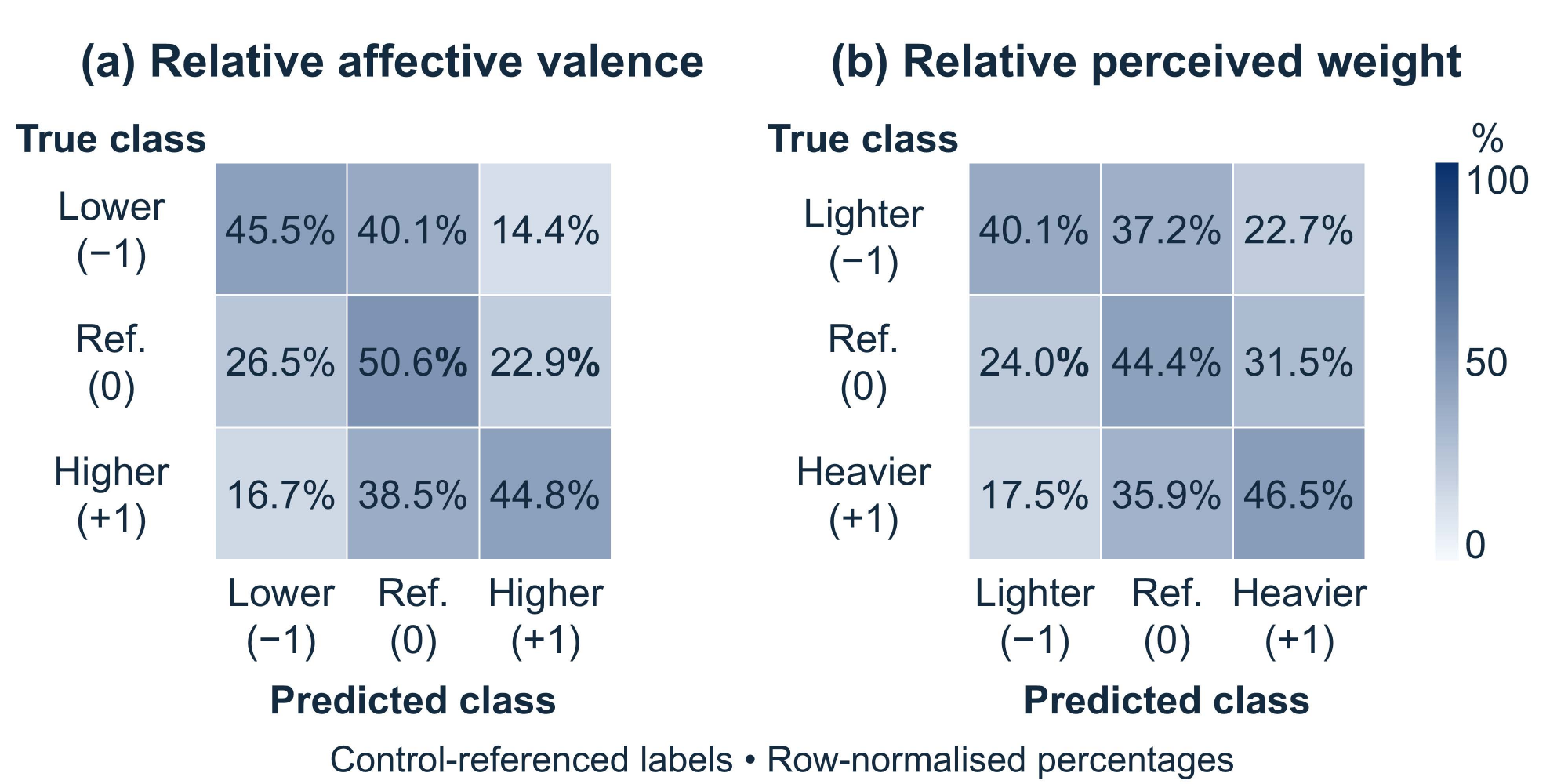}
    \caption{Confusion matrices for relative (a) affective valence and (b) perceived weight. Window-level predictions and targets were pooled across participants and 10 random seeds. Ref. (0) includes control trials and manipulated-feedback trials rated equal to the participant-specific control mean. Negative/positive labels indicate lower/higher valence or lighter/heavier perceived weight relative to that mean.}
    \label{fig:confusion_matrices}
\end{figure}

Overall accuracy is the proportion of correctly classified test windows. To summarise BA, macro F1, and overall accuracy across participants, we averaged each participant's scores over the 10 random seeds and then calculated the mean and standard deviation of these participant-level averages. This gives each participant equal weight in the reported metrics. For descriptive comparison, we also report the proportion of participants whose mean BA exceeded the nominal three-class reference of 33.3\%. Figure~\ref{fig:confusion_matrices} provides the confusion matrices for the subjective tasks, showing predictions pooled across participants and seeds.

\subsection{Benchmark Results}
\label{sec:results}

Table~\ref{tab:classification_results} summarises classification performance and task-specific participant counts. Auditory-condition classification achieved a mean balanced accuracy of 36.6\%, compared with the nominal three-class reference of 33.3\%. Mean balanced accuracy was higher for relative perceived weight ($44.5 \pm 11.5$\%) and relative affective valence ($45.9 \pm 12.7$\%). The reported $\pm$ values are the sample standard deviations across participants after averaging each participant's balanced accuracy over the 10 random seeds. For the auditory condition, 61.6\% of participants had balanced accuracy greater than 33.3\%, 81.5\% for relative perceived weight, and 84.3\% for relative affective valence.

\begin{table*}
    \caption{Within-participant classification performance across eligible participants and 10 random seeds. $N$/pool denotes the number of eligible participants relative to the task-specific pool. Class distributions are reported as average participant-level percentages, with subjective-task class distributions ordered as $[-1, 0, +1]$. $>$ref. denotes the percentage of participants with balanced accuracy exceeding the nominal three-class reference value of 33.3\%. All performance values are percentages.}
    \label{tab:classification_results}
    \centering
    \small
    \setlength{\tabcolsep}{4pt}
    \renewcommand{\arraystretch}{1.1}
    \begin{tabularx}{0.7\textwidth}{@{}
        >{\raggedright\arraybackslash}X
        c c c c c c@{}}
        \toprule
        \textbf{Task} &
        \boldmath $N/\text{pool}$ &
        \textbf{Distribution} &
        \textbf{Accuracy} &
        \textbf{BA} &
        \textbf{F1} &
        \textbf{$>$ref.} \\
        \midrule
        Auditory condition &
        99/99 & 34.4/33.8/31.8 &
        34.8 & 36.6 & 28.5 & 61.6 \\
        \rowcolor[HTML]{EFEFEF}
        Relative perceived weight &
        92/98 & 23.3/41.4/35.3 &
        43.9 & 44.5 & 37.6 & 81.5 \\
        Relative valence &
        83/98 & 26.5/47.0/26.5 &
        47.9 & 45.9 & 39.4 & 84.3 \\
        \bottomrule
    \end{tabularx}
\end{table*}

\begin{figure*}
    \centering
    \subfigure[]{
        \includegraphics[width=0.48\linewidth]{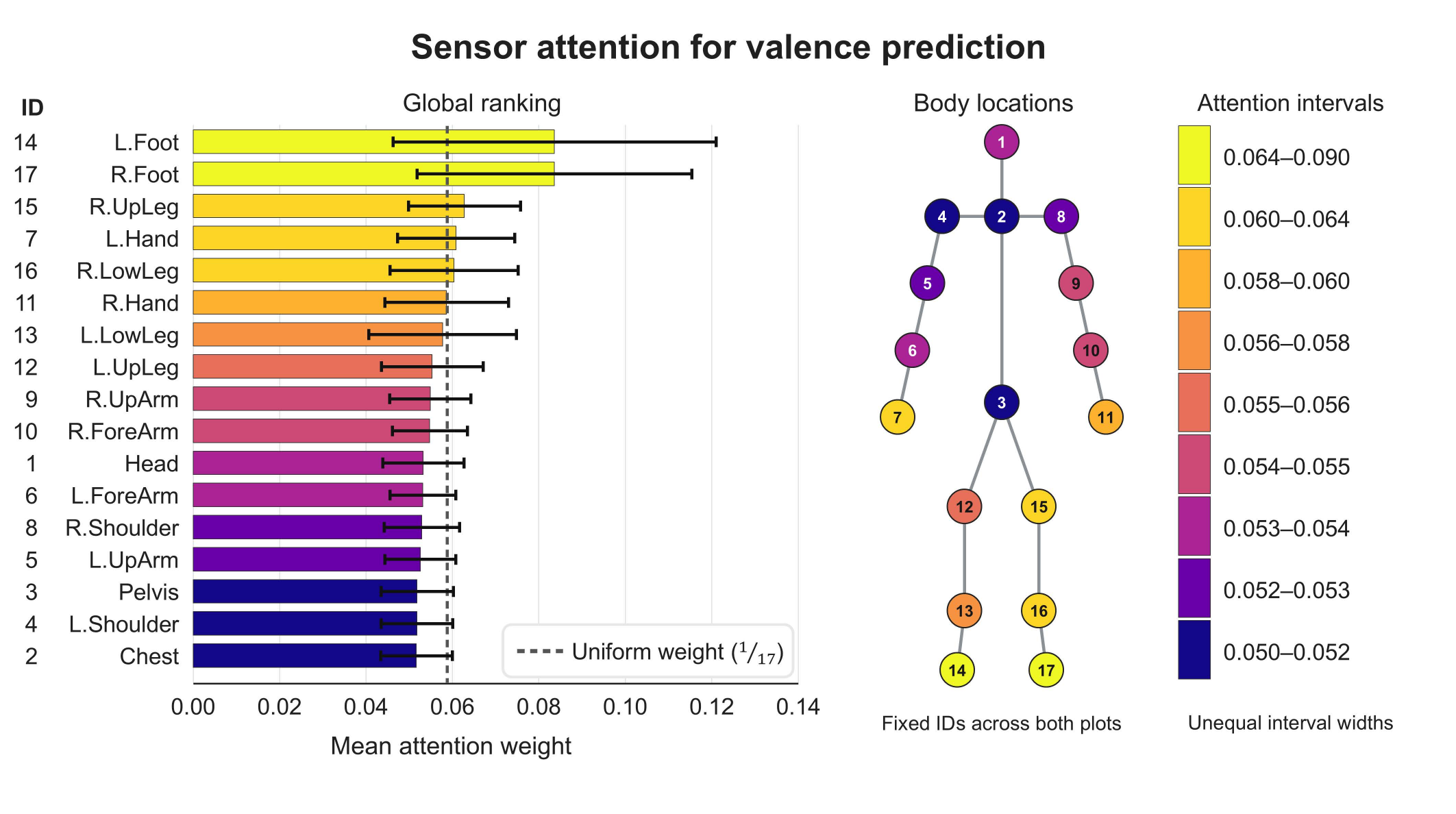}
        \label{fig:first_sub}
    }
    \subfigure[]{
        \includegraphics[width=0.48\linewidth]{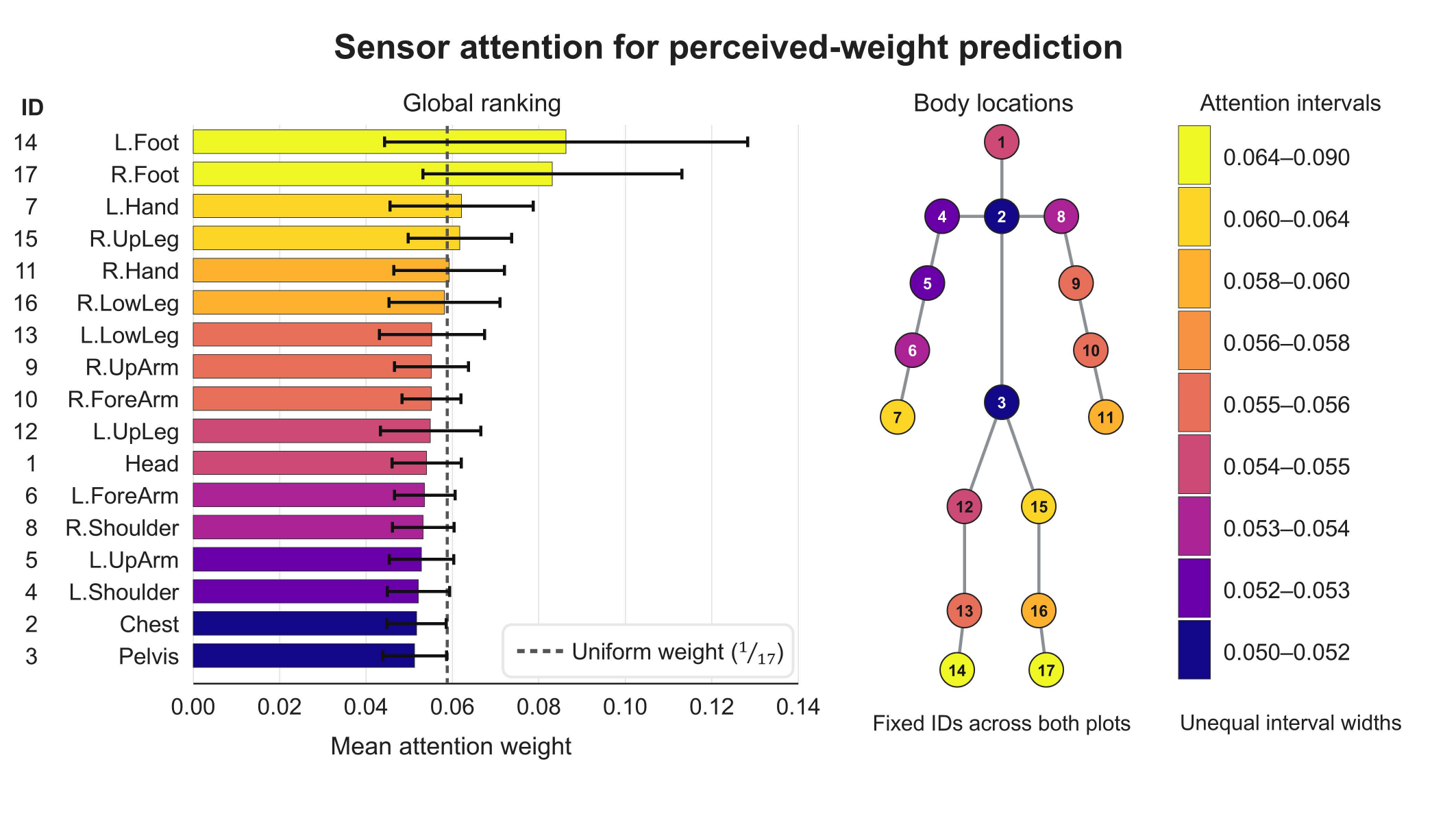}
        \label{fig:second_sub}
    }
    \caption{Attention profiles for (a) relative affective valence and (b) relative perceived weight. Weights were averaged within participants across test windows and then across 10 seeds. Bars show participant-level means averaged across participants, and error bars show population standard deviations across participants.}
    \label{fig:sensor_importance_subjective}
\end{figure*}

Alongside classification performance, the subjective-task attention profiles describe how the benchmark model allocates weights across body locations. For each participant, we averaged attention weights across test windows within each seed and then across the 10 seeds. Figure~\ref{fig:sensor_importance_subjective} presents the attention profiles for the two subjective tasks. The models assigned the largest average weights to the foot sensors, followed by contributions from the hands and upper legs. Trunk and proximal locations remained closer to the uniform-weight reference. The two profiles were broadly similar, with differences mainly in the relative weighting of individual locations.

Error bars overlapped for many locations with intermediate weights. The profiles therefore provide a broad description of weight allocation rather than a precise ranking of body regions.

\section{Discussion}
\label{sec:discussion}

BODYinTRANSIT-Lab connects controlled auditory feedback with multimodal recordings, complementary subjective annotations, and participant characteristics. The benchmark examines one computational use of this resource: classifying auditory conditions and relative post-trial reports from walking kinematics. We discuss what the dataset and these results support, how the observations may inform further work, and the methodological and ethical considerations that accompany the computational study of BTEs.

\subsection{Complementary Annotations and Further Computational Uses}

The dataset supports investigation of relationships among sensory manipulation, reported experience, and bodily signals without requiring these measures to be interchangeable. Seven perceived-weight reports per trial provide repeated observations of perceived weight, while post-trial questionnaires, avatar estimates, and body maps capture affective, scalar, visual, and spatial aspects of the experience. These complementary annotations support investigation of how different aspects of a BTE relate to one another and, through the repeated within-trial reports, enable fine-grained analysis of how the BTE emerges and unfolds over time \cite{turmovidal_trajectories2025}. Participant-level measures add context for examining variation between individuals. These observations also allow researchers to study both convergence and divergence among responses to the same feedback.

The present benchmark uses only acceleration signals and two post-trial rating measures. Further analyses could examine relationships between reported body size and felt weight, and associations between participant characteristics and response patterns. Combining kinematics with ECG, EDA, EMG, and respiration may provide complementary information for these analyses. The dataset's availability, along with its acquisition and curation protocol, supports secondary analyses and the development of comparable resources without repeating the entire experimental procedure~\cite{Roel_pos_2026,open_science_Gilmore_Diaz_Wyble_Yarkoni_2017,open_science_Hales_Wesselmann_Hilgard_2019}.

\subsection{Interpretation of the Kinematic Benchmark}

The subjective tasks achieved higher mean balanced accuracy than auditory-condition classification under the reported evaluation protocol. This pattern aligns with the distinction established in the study design: the delivered auditory feedback and the experience reported under it are related but target different constructs. These results provide initial reference values for predicting control-referenced perceived weight and affective valence labels from walking kinematics within recorded sessions. Because training and test subsets may contain segments from the same trial, the evaluation does not distinguish information associated with reported experience from other trial-specific movement characteristics. They do not establish that subjective experience mediates an effect of auditory feedback on movement, or that condition-related information has been eliminated from the subjective tasks.

Interpretation also depends on how the labels are constructed. Both control trials were assigned class 0, while ratings under HF and LF feedback were compared with the participant's control mean. The subjective targets therefore combine the experimental reference with the relative rating. Moreover, the three tasks differ in class distributions and eligible participant pools. Their performance contrast should therefore be treated as a descriptive comparison under the specified benchmark, rather than a direct comparison of the strengths of perceptual, affective, and motor effects.

The benchmark model provides a baseline for combining pretrained movement representations across body locations. The attention profiles place the largest average weights on the feet for both subjective tasks, motivating evaluation of whether a smaller sensor configuration could retain useful predictive information. This possibility requires direct comparisons of sensor subsets, supported by systematic sensor ablation~\cite{hens2026one}. Similarly, the contribution of attention pooling to artefact robustness requires comparison with alternative aggregation methods; the present results alone do not isolate that contribution.

\subsection{Ethical Considerations for BTE Computing}

Although automatic detection of BTEs has the potential to support interventions for health and well-being, it also raises important ethical challenges. Body-perception technologies occupy a particularly sensitive position because they both influence bodily experience and infer internal psychological states. As recently argued in HCI research~\cite{floyd_grandchallenges}, such systems require careful consideration of agency, informed consent, transparency, and trust, particularly when technologies adapt their behaviour in real time or influence users outside their conscious awareness. Moreover, physiological and behavioural signals should not be treated as objective proxies for subjective experience. Body perception and emotion are inherently contextual, culturally situated, and individually variable, making algorithmic overconfidence and inappropriate personalisation significant risks.

In this regard, relying on biometrics, movement sensing, and computational models risks reducing rich, lived bodily experiences to quantitative metrics or categorical labels. Such reductions become particularly problematic when adaptive systems aim to guide users toward a ``preferred" bodily or emotional state, raising questions about who defines these states and according to which values. Because BTEs inevitably intersect with cultural norms surrounding appearance, ability, gender, ageing, and emotional expression, computational models and feedback mechanisms may inadvertently reinforce normative assumptions about what constitutes a ``healthy," ``proper," or ``ideal" body~\cite{tajadurajimenez_2026_CHI_panel}.

Addressing these challenges requires human-centred approaches that make underlying assumptions explicit, recognise that sensing and adaptation technologies are not value-neutral, and preserve users' autonomy throughout the interaction~\cite{tajadurajimenez_2026_CHI_panel}. Future systems should enable users to understand and challenge system inferences, minimise the unnecessary collection of sensitive physiological data, evaluate fairness across diverse populations~\cite{floyd_grandchallenges}, as well as support opportunities for users to define meaningful BTEs on their own terms rather than imposing normative models of embodiment~\cite{tajadurajimenez_2026_CHI_panel}. We therefore argue that ethical reflection should be treated as a core design principle of computational models of BTEs from the outset, particularly as these technologies move from laboratory settings into everyday and clinical environments.

\subsection{Limitations and Future Work}

The benchmark evaluates held-out walking segments from participants whose other segments are available for training. Because partitioning occurs at segment level, observations from the same trial may appear in different subsets. Performance therefore characterises within-session prediction and does not establish generalisation to unseen trials, sessions, or participants. Evaluation that holds out complete trials, followed by separate-session and participant-independent studies, would address progressively broader prediction settings.

The subjective tasks also depend on a reference estimated from two control ratings. This reference is sensitive to variation in those ratings, and the resulting classes retain direction rather than the magnitude of the difference. Future benchmarks could compare alternative reference and target constructions, including regression, and evaluate their stability.

Annotation timing places an additional constraint on the conclusions. The seven within-trial perceived-weight ratings are repeated discrete observations, while post-trial ratings provide one assessment per trial. Assigning a post-trial label to walking windows does not establish the participant's state at every moment within those windows. More frequent or continuous annotation could support investigation of temporal changes~\cite{Finotti_2023_continuous_measure}, but its use during walking must account for the movement and cognitive demands introduced by reporting. The distinction between repeated and post-trial annotation therefore remains relevant even when the underlying sensor signals are continuous.

The sample consisted predominantly of young adults recruited in Madrid, Spain. Replication across age groups, cultural contexts, and health conditions would help assess the transferability of observations and models~\cite{hofstede2011cultures,Swami2021}. The protocol also concerns short-term exposure to auditory feedback. Longitudinal studies would be needed to examine the persistence of perceptual changes and their relationship to longer-term changes in the physical body~\cite{Weijs_Cortex}. Adaptive, everyday, or clinical uses would additionally require prospective evaluation of prediction reliability and the effects of acting on model outputs.

\section{Conclusion}

BODYinTRANSIT-Lab provides a multimodal dataset for investigating Body Transformation Experiences (BTEs) through the Footsteps Illusion, in which participants hear their own footstep sounds modified in real time as they walk. Recordings from 101 adults, recruited through stratification by eating-disorder symptomatology and physical activity, link repeated auditory conditions to full-body kinematics, physiology, and subjective assessments. This design allows responses to the same feedback to be examined within individuals and in relation to differences in their characteristics.

Seven within-trial perceived-weight ratings capture variation during walking, whereas post-trial bodily and affective assessments, avatar estimates, and body maps describe additional perceptual and spatial aspects of the experience. Linking these observations to bodily recordings supports investigation of how movement and physiology relate to different aspects of reported experience.

As an initial computational benchmark, we proposed a deep learning architecture that combines pretrained movement representations with attention-based pooling across body locations. Using walking acceleration, it achieved mean balanced accuracies of 36.6\% for the auditory condition, 44.5\% for relative perceived weight, and 45.9\% for relative affective valence when evaluated on held-out walking segments within participants' recorded sessions.

Beyond this kinematic benchmark, the physiological recordings support an investigation into whether combining movement and physiology improves predictions of reported bodily and affective experiences. The accompanying acquisition and curation protocol documents how we collected, synchronised, prepared, and organised the observations to provide a basis for other researchers to construct comparable multimodal datasets for studying BTEs. Together, the dataset and documented protocol provide a reusable basis for investigating variation in bodily and subjective responses within and across individuals and for collecting comparable multimodal data on BTEs.

\section{Data Availability}
The BODYinTRANSIT-Lab database is made available to the research community to promote reproducible research. The proposed benchmark implementation is available at~\url{https://github.com/tmcortes/BODYinTRANSIT}. Due to the sensitive nature of the data, access is provided under specific conditions. Interested researchers are required to complete and sign an End-User Licence Agreement. Upon verification, researchers will be provided with credentials to access the dataset ($\sim$15 GB) from a secure repository at the Universidad Carlos III de Madrid.

\section*{Acknowledgement}
This work was supported by the European Union's Horizon 2020 program under grant No. 101002711 for the BODYinTRANSIT project.

\appendices
\section{Participant Demographics}
Table~\ref{tab:demographics} shows Demographics, Anthropometric, Psychometric, and Social Characteristics for 101 participants.

\renewcommand{\thetable}{S\arabic{table}}
\setcounter{table}{0}
\begin{table*}
    \caption{Participant Demographics, Anthropometric, Psychometric, and Social Characteristics (N=101). \textbf{Sex (F/M)}: Sex assigned at birth (Woman / Man); \textbf{Gender (F/M/NB/O)}: Self-reported gender (Feminine / Masculine / Non-binary / Other or Prefer not to say); \textbf{Education Level (1-5)}: Ordinal education level (1=Secondary, 2=Vocational, 3=Bachelor's, 4=Master's, 5=PhD); \textbf{Gender Identity (1-7)}: Self-perceived gender identity (1=very masculine, 7=very feminine); \textbf{Ideal Gender Identity (1-7)}: Desired gender identity (1=very masculine, 7=very feminine); \textbf{EDE-Q}: Eating Disorder Examination Questionnaire (Global Score); range 0--6, where 0 indicates no/lowest frequency of eating-disorder-related symptoms and 6 indicates the highest frequency/severity of symptoms; \textbf{IPAQ}: International Physical Activity Questionnaire (total score in MET-min/week); quantifies physical activity based on reported activity during the previous 7 days, expressed in Metabolic Equivalent of Task (MET) minutes per week; \textbf{MBSRQ}: Multidimensional Body-Self Relations Questionnaire (MBSRQ) subscale scores: Importance of Corporality (ISC), Behaviours aimed at maintaining physical shape (COMF), Self-evaluated physical attractiveness (AFA), and Physical appearance care (CAF); range 1--5, where higher scores indicate greater endorsement of the corresponding dimension; \textbf{Betts' QMI}: Overall Imagery Vividness score from the Betts' Questionnaire upon Mental Imagery; range 1--7, where a higher numerical score actually indicates lower imagery vividness; \textbf{SSQ6}: Statistics describing the participant's self-reported social network, as measured by the Social Support Questionnaire (Sex Ratio, Network Size, Group Count, Average Age of network, Age Difference with participant, and Mean \& Standard deviation (SD) of people per questionnaire item); assesses perceived social support, including the number of available supportive people and satisfaction with that support, where higher scores indicate greater perceived availability of social support. Group 1: Low EDE-Q / Low IPAQ, Group 2: High EDE-Q / Low IPAQ, Group 3: Low EDE-Q / High IPAQ, Group 4: High EDE-Q / High IPAQ.}
    \label{tab:demographics}
    \centering
    \begin{tabular}{lccccc}
        \toprule
        \textbf{Characteristic} & \textbf{Group 1 (n=20)} & \textbf{Group 2 (n=21)} & \textbf{Group 3 (n=41)} & \textbf{Group 4 (n=19)} & \textbf{Overall (N=101)} \\
        \midrule
        \textbf{Age (years)} & 24.25 (8.72) & 25.81 (7.38) & 25.10 (6.50) & 25.11 (7.85) & 25.08 (7.32) \\
        \textbf{Sex (F/M)} & 16 / 4 & 11 / 10 & 19 / 22 & 8 / 11 & 54 / 47 \\
        \textbf{Gender (F/M/NB/O)} & 16 / 3 / 0 / 1 & 12 / 9 / 0 / 0 & 18 / 22 / 1 / 0 & 8 / 10 / 1 / 0 & 54 / 44 / 2 / 1 \\
        \addlinespace
        \textbf{Height (cm)} & 169.55 (9.03) & 168.52 (8.30) & 171.32 (8.49) & 169.58 (7.55) & 170.06 (8.35) \\
        \textbf{Weight (kg)} & 65.35 (12.21) & 65.76 (12.47) & 67.85 (10.61) & 62.37 (5.86) & 65.89 (10.71) \\
        \textbf{Ideal Weight (kg)} & 62.06 (11.82) & 62.67 (9.17) & 65.95 (9.93) & 62.76 (6.43) & 63.96 (9.65) \\
        \textbf{Waist (cm)} & 75.35 (9.17) & 78.33 (9.37) & 78.05 (11.26) & 73.47 (6.11) & 76.71 (9.72) \\
        \textbf{Hip (cm)} & 91.70 (8.52) & 92.19 (14.73) & 93.51 (9.73) & 90.79 (8.45) & 92.37 (10.44) \\
        \textbf{Shoe Size (EU)} & 39.30 (2.39) & 40.43 (2.98) & 40.41 (2.52) & 39.47 (3.91) & 40.02 (2.90) \\
        \addlinespace
        \textbf{Education Level (1-5)} & 1.95 (1.36) & 2.33 (1.49) & 2.10 (1.28) & 2.21 (1.18) & 2.14 (1.31) \\
        \textbf{Gender Identity (1-7)} & 5.16 (1.89) & 4.48 (1.83) & 3.95 (1.80) & 4.68 (1.92) & 4.43 (1.87) \\
        \textbf{Ideal Gender Identity (1-7)} & 5.05 (1.75) & 4.19 (1.83) & 3.95 (1.80) & 4.58 (2.04) & 4.33 (1.86) \\
        \addlinespace
        \textbf{EDE-Q (0-6)} & 0.55 (0.44) & 2.00 (0.82) & 0.60 (0.41) & 2.01 (0.66) & 1.15 (0.90) \\
        \textbf{IPAQ (MET-min/week)} & 1802 (773) & 2380 (1479) & 7249 (5872) & 6881 (5953) & 5089 (5191) \\
        \textbf{MBSRQ ISC (1-5)} & 1.53 (0.37) & 1.42 (0.37) & 1.57 (0.26) & 1.71 (0.33) & 1.56 (0.33) \\
        \textbf{MBSRQ COMF (1-5)} & 1.26 (0.35) & 1.59 (0.30) & 1.56 (0.28) & 1.63 (0.26) & 1.52 (0.32) \\
        \textbf{MBSRQ AFA (1-5)} & 4.33 (0.75) & 4.21 (1.02) & 4.35 (0.85) & 4.30 (0.91) & 4.31 (0.87) \\
        \textbf{MBSRQ CAF (1-5)} & 4.68 (0.63) & 4.28 (0.71) & 4.60 (0.76) & 4.77 (0.44) & 4.58 (0.69) \\
        \textbf{Betts' QMI (1-7)} & 2.68 (1.25) & 2.82 (0.70) & 2.40 (0.63) & 2.61 (0.74) & 2.58 (0.83) \\
        \addlinespace
        \textbf{SSQ6: Sex Ratio} & 0.60 (0.26) & 0.54 (0.19) & 0.49 (0.24) & 0.48 (0.21) & 0.52 (0.23) \\
        \textbf{SSQ6: Network Size} & 4.80 (2.24) & 4.86 (1.88) & 4.66 (2.46) & 4.32 (1.80) & 4.66 (2.16) \\
        \textbf{SSQ6: Groups Count} & 1.25 (0.55) & 1.81 (0.87) & 2.02 (1.21) & 1.95 (0.97) & 1.81 (1.03) \\
        \textbf{SSQ6: Average Age} & 30.96 (12.35) & 36.29 (8.37) & 34.50 (7.98) & 32.48 (10.01) & 33.86 (9.37) \\
        \textbf{SSQ6: Age Difference} & 6.23 (7.13) & 10.07 (8.95) & 9.52 (8.95) & 7.36 (7.52) & 8.63 (8.37) \\
        \textbf{SSQ6: People/Item (Mean)} & 3.03 (1.25) & 3.02 (0.96) & 3.00 (1.38) & 2.85 (0.76) & 2.98 (1.16) \\
        \textbf{SSQ6: People/Item (SD)} & 1.16 (0.70) & 1.20 (0.62) & 1.04 (0.62) & 1.11 (0.57) & 1.11 (0.62) \\
        \bottomrule
    \end{tabular}
\end{table*}

\bibliographystyle{IEEEtran}
\bibliography{references}

\end{document}